\documentclass[useAMS, usenatbib]{mn2e}
   
\usepackage{epsfig}%,natbib}

\newcommand{\kms}{\mathrm{\;km\;s^{-1}}}
\newcommand{\mN}{\mathcal{N}}

\title[Blue fraction evolution in DEEP2 groups]{The DEEP2 Galaxy
  Redshift Survey: The evolution of the blue 
fraction in groups and the field}

\author[Brian F. Gerke et al.]{Brian
  F. Gerke$^{1}$\thanks{E-mail:bgerke@astro.berkeley.edu}, 
Jeffrey A. Newman$^{2}$, S. M. Faber$^{3}$, Michael
C. Cooper$^{4}$,\newauthor 
Darren J. Croton$^{4}$, Marc Davis$^{1,4}$, Christopher
N. A. Willmer$^{5}$,
Renbin Yan$^{4}$, \newauthor 
Alison L. Coil$^{5}$, Puragra Guhathakurta$^{3}$, David
C. Koo$^{3}$, Benjamin J. Weiner$^{6}$\\
$^{1}$Department of Physics, University of California,
Berkeley, CA 94720\\
$^{2}$Hubble Fellow; Institute for
Nuclear and Particle Astrophysics, Lawrence Berkeley National Laboratory, 
Berkeley, CA 94720 \\
$^{3}$University of California Observatories/Lick
Observatory, Department of Astronomy and Astrophysics, University of
California, \\ Santa Cruz, CA 95064\\
$^{4}$Department of Astronomy, University of California,
Berkeley, CA 94720\\
$^{5}$Steward Observatory, University of Arizona, Tucson,
  AZ 85721\\
$^{6}${Department of Astronomy, University of Maryland,
College Park, MD  20742}}
\begin{document}
\date{\today}
\maketitle

\label{firstpage}

\begin{abstract}

We explore the behavior of the blue galaxy fraction over the redshift
range $0.75\le z\le 1.3$ in the DEEP2 Survey, both for field galaxies
and for galaxies in groups. 
The primary aim 
is to determine the role that groups play in driving the
evolution of galaxy colour at high $z$.  In pursuing this aim, it is
essential to define a galaxy sample that does not
suffer from redshift-dependent selection effects in colour-magnitude
space.  We develop four such samples for this study: at all redshifts
considered, each one is complete in colour-magnitude space, and the
selection also accounts for evolution in the galaxy luminosity
function.  These samples will also
be useful for future evolutionary studies in DEEP2.  The colour
segregation observed between local group and field samples is already
in place at $z\sim 1$: DEEP2 groups have a significantly lower blue 
fraction than the field.  At fixed $z$, there is also a correlation
between blue fraction and galaxy magnitude, such that brighter
galaxies are more likely to be red, both in groups and in the
field.  In addition, there is a negative correlation 
between blue fraction and group richness.  In
terms of evolution, the blue fraction in groups and the field remains
roughly constant from $z=0.75$ to $z\sim 1$, but beyond this redshift
the blue fraction in groups rises rapidly with $z$, and
the group and field blue fractions become indistinguishable at $z\sim
1.3$.  Careful tests indicate that this effect does not arise from
known systematic or selection effects.  To further ensure the
robustness of this result, we build on previous mock DEEP2 catalogues
to develop mock catalogues that reproduce the
colour-overdensity relation observed in DEEP2 and use these to test
our methods.  The convergence between the group and field blue
fractions at $z\sim 1.3$ implies that DEEP2 galaxy groups only became
efficient at 
quenching star formation at $z\sim 2$; this result is
broadly consistent with other recent observations and with current
models of galaxy evolution and hierarchical structure growth.

\end{abstract}


%%%%%%%%%%INTRODUCTION%%%%%%%%%%%%%%%%%%%%%%%%%%%%%%

\section{Introduction}
\label{sec:intro}
One of the most striking characteristics of the galaxy
population is the well-known environmental segregation of the two main
galaxy types: red, early-type galaxies with little ongoing star
formation preferentially occur in galaxy groups and clusters, while
blue, late-type galaxies with significant star-formation activity
avoid such systems and preferentially populate the
``field''\footnote{Throughout this paper, we shall use the term
  \emph{field} to refer to those galaxies that are not in groups
  (i.e., isolated galaxies), not the full galaxy population}.  This
observation has been recognized as a key to understanding galaxy
formation and evolution for more than fifty years \citep{SB51}.%, and 

There is now overwhelming evidence that the galaxy population in
clusters has evolved significantly with redshift down to the present
day.  \citet{BO84} were the first to present evidence that the
fraction of blue galaxies, $f_b$, in clusters increases strongly with
increasing $z$---the so-called Butcher-Oemler (BO) effect.  This basic
result---an increased incidence of star-forming galaxies in
distant clusters---has been 
replicated in numerous later studies using a variety of star-formation
indicators, including cluster blue fractions
\citep{RS95, MD00, KB01, Ellingson01, Margoniner01},  emission-line
galaxy fractions \citep{Poggianti06}, and morphological fractions
\citep{ODB97, Couch98, vanD00, Fasano00,
  LOP02, Goto03}.  Such studies have also been extended to less massive
galaxy groups 
\citep{AS93, Wilman05b, MOL06}, which also appear to have
proportionally more star-forming galaxies   
back in time. To be sure, there remain strong reasons to question
the veracity of the BO effect as it was \emph{originally} presented
for  
galaxies in the \emph{cores} of \emph{rich} clusters (\emph{e.g.},
\citealt{Koo88, 
  Smail98, Andreon06b}).  But it is now indisputable that 
clusters, on the whole, had proportionally more blue, star-forming,
and morphologically late-type 
galaxies---in short, more star formation---at $z\sim 0.5$
than they do at present.  

It is tempting to conclude that this
evolution is responsible for the substantial growth that has been
observed in the number density of red galaxies since $z\sim 1$
\citep{Bell04b, Willmer06, Faber06}. But it is also important to note
that, in addition to evolution \emph{within} the group and cluster
environment, the 
build-up of the red galaxy population could also be driven by an
\emph{increasing number density} of groups and clusters from the
growth of 
structure. We shall try in this paper to shed light on the relative
importance in DEEP2 of these two channels for red-galaxy formation.  

%The BO effect
A blue fraction that declines with time in groups and clusters is a
natural consequence of  
hierarchical schemes for galaxy formation \citep{Kauffmann95a, BFC96, 
  Diaferio01, Benson01} if one assumes that 
groups and clusters play a role in quenching star formation.  Since
these systems form at  
relatively late times, their member galaxies at intermediate redshift
will have had less time, on average, to
``feel'' the effects of group membership; thus, a smaller fraction of
them will have ceased forming stars.  Indeed, it has been shown
\citep{KB01} that the evolution of $f_b$ in clusters out to $z\sim
0.4$ can be entirely accounted for by considering the effects of
hierarchical structure formation and a declining universal
star-formation rate.  For exactness in what
follows, then, we shall use the phrase ``Butcher-Oemler effect'' to
refer just to this effect---a 
decline in the typical blue fraction in groups and clusters that
results \emph{simply} from the  
increasing age of these systems with time, and \emph{not} from any
evolution in the quenching efficiency of groups and clusters.  
Although this working definition differs somewhat from the effect
originally 
reported by \citet{BO84}, it reflects usage that has become common
in recent literature.  It is also worth noting that, under this
definition, the BO effect may be stronger in some types of systems
than in others (in clusters than in groups, for example), depending on  
the accretion rates and quenching efficiencies involved.

It remains unclear,
however, which specific physical mechanisms are responsible for
quenching star formation, over what timescales they act, and even
whether these mechanisms are peculiar to groups and clusters.   
A variety of mechanisms %for quenching in overdense environments 
has
been  proposed, including (\emph{1}) galaxy mergers (\emph{e.g}, 
\citealt{TT72}), which occur primarily in galaxy groups
(\emph{e.g.}, \citealt{CCM92}) and can trigger AGN feedback that
sweeps merger remnants of 
their gas (\citealt{SDH05}); (\emph{2}) close,
high-velocity 
galaxy encounters (``harrassment''; \citealt{Moore96}), which occur
primarily in clusters; (\emph{3}) ram-pressure stripping of galaxy
gas by the hot intracluster medium (\citealt{GG72}); and
(\emph{4}) the less violent process known as ``strangulation'' in
 which gas accretion onto galaxy discs is cut off, either by stripping of
galaxies' gaseous haloes (\citealt{LTC80}) or by feedback
from low-luminosity AGN \citep{Croton06},
and star formation ends when 
the remaining supply of gas has been exhausted.  There is evidence
that each 
of these processes is at work on some level; it may be that no
single mechanism bears primary responsibility for galaxy evolution.
Disentangling how and to what degree each of the above processes shapes
the galaxy population requires detailed observations of galaxies over
a wide range of redshifts and environments.

The observed evolution is rather complicated in its details, however.
For example, the well-known morphology-density relation for galaxies
in clusters (\emph{e.g.},  \citealt{Dressler80}) and groups
(\emph{e.g.} \citealt{PG84}) evolves strongly with
increasing redshift \citep{Dressler97}, and there is evidence that the 
evolution in intermediate-density environments has occured more
recently than in high-density environments \citep{Smith05}. It also
appears that the morphological mix of cluster early-types has changed
with time \citep{Fasano00, Postman05}.  Complicating things further,
cluster blue fractions display a substantial
system-to-system scatter at all epochs \citep{BO84, Smail98, MD00,
  Goto03, DeP04}.  This scatter can be ascribed in part to the
observed trends between blue fraction and cluster mass,
\emph{i.e.}, velocity dispersion \citep{NKB88}, inferred virial mass 
\citep{Martinez02}, richness \citep{Margoniner01, Goto03, TPA05}, or
total 
optical  
luminosity \citep{Weinmann06}; the correlation between $f_b$ and mass 
also appears to evolve with redshift \citep{Poggianti06}.  The scatter
may also arise in part from a correlation between $f_b$ and a cluster's
degree of dynamical relaxation (\emph{e.g.}, \citealt{MRU00}).   The
scatter in $f_b$ 
might mask or enhance the redshift 
evolution, depending on the cluster selection criteria used
\citep{NKB88, AE99}.  Indeed, \citet{Smail98} observe no
BO effect out to $z\sim 0.25$ in a sample of luminous
X-ray clusters (but see also \citealt{Fairley02}), and many authors
have commented on the presence of individual massive clusters at high
redshift with low $f_b$ and well-formed red sequences (\emph{e.g.}
\citealt{Koo81, Homeier05, Andreon06}).  It is clear that a robust
study of galaxy evolution in groups and clusters requires a sample
that contains a broad range of systems and is uniformly
selected at all redshifts.   

Until relatively recently, studies of galaxy evolution in clusters
were limited to cluster samples pre-selected from optical imaging or
X-ray maps; such samples are prone to selection effects that bias the
galaxy sample as a function of $z$.  The advent of large, densely
sampled galaxy 
redshift surveys like the 2-degree Field Galaxy Redshift Survey
(2dFGRS) and the Sloan Digital Sky Survey (SDSS),
however, has allowed selection of large, low-redshift cluster samples
directly from 
the galaxy distribution in redshift space, where redshift-dependent
selection effects can be well understood and accounted for.  Several
authors have studied the properties of the cluster galaxy population
in the 2dFGRS \citep{Balogh04, DeP04} and SDSS \citep{Goto05a, Goto05b,
  Weinmann06, Quintero06}.  Within SDSS, there have even 
been detections of an evolving $f_b$ in clusters \citep{Goto03,
  MOL06} and of an 
evolving morphology-density relation \citep{GYTO04}.  With the
high-redshift DEEP2 Galaxy  
Redshift Survey (\citealt{DGN04}, Faber et al. in prep.)
nearing completion, it is now possible to extend 
these studies to $z\sim 1$.

DEEP2 has now yielded 
a catalogue of several thousand groups and small
clusters over a wide range of masses at $z\sim 1$ \citep{Gerke05}.  A
study of 
the DEEP2 sample by \citet{Cooper06a} has already revealed that,
\emph{without} explicitly considering groups,  the
correlations between galaxy properties and the density of the
nearby galaxy distribution are qualitatively similar at $z\sim 1$ to
what is observed locally (\emph{e.g.},  \citealt{Balogh04, Hogg04}),
although the relations differ in detail.  Interestingly,
recent studies have 
shown that these correlations evolve with redshift, becoming
stronger 
over time \citep{Nuijten05, Cucciati06, Cooper06b}, and that the
color-density relation in the \emph{local} 
Universe can be ascribed almost entirely to mechanisms acting in
groups and clusters \citep{BBH06}.   
The primary goal of this paper, then, is  to establish the role that
groups and clusters play, if any, in driving the evolution of the
DEEP2 galaxy population.   In 
particular, we shall explore the possibility that 
group and cluster environments are responsible for the build-up of red
galaxies observed over the DEEP2 redshift range \citep{Bell04b,
  Willmer06, Faber06}; by so doing, we 
will try to shed light on the physical mechanisms driving the evolution.  
Although this study will focus on the evolutionary effects of
galaxies' membership in groups and clusters (that is to say, of their
dark matter halo masses), 
there is important complementary information to be gained by studying 
the effects of the local density of galaxies.  Such a study is
undertaken in a companion paper to this one by \citet{Cooper06b}. 
For now, however, we
will measure $f_b$  for galaxies in groups and
clusters, as 
a function of redshift, and compare it to that found in the general field
(\emph{i.e.}, in the galaxy population outside of groups).  We will
also investigate the relation between $f_b$ and cluster
properties in order to gain insight into the processes at work and
also to understand and control any systematic selection effects.

The thoughtful reader may find it perverse to use the
\emph{blue} fraction to study the evolution of \emph{red} galaxies,
but $f_b$ has significant historical precedent in the study of cluster 
galaxy populations, so we use it for consistency with earlier studies.
In any case, the 
possibly more natural \emph{red} fraction statistic is simply $f_r =
1-f_b$.  We have chosen
to consider galaxy \emph{colour} rather than 
other indicators of galaxy type like morphology or [O II] line
emission primarily because it can be measured accurately for the
largest number of DEEP2 galaxies, allowing for robust statistics.  The
DEEP2 
ground-based imaging lacks the resolution necessary for morphological
classification of most galaxies at $z\sim 1$, and there is 
\emph{HST} imaging for only a small fraction of the sample.
Emission line strength can be measured accurately only in spectra with
sufficient signal-to-noise ratio; not all DEEP2 galaxies meet this
criterion.  Also, it has recently been shown that [O II] emission
may not give a clear indication of star formation, especially in red
galaxies \citep{Yan06}. 
Regardless, there is a strong and relatively tight correlation between
[O II] equivalent width and galaxy colour% in DEEP2
(\emph{e.g.}, \citealt{Weiner05, Cooper06a}), so the two statistics
should give similar results.

We shall proceed as follows.  In \S \ref{sec:DEEP2}, we describe the
DEEP2 survey and the DEEP2 group catalogue, and in \S \ref{sec:mocks} we
discuss the construction of mock DEEP2 galaxy 
catalogues appropriate for this study.  We describe our galaxy selection
criteria and incompleteness corrections and present the precise
definition of $f_b$ used here in \S \ref{sec:selection}. 
Measurements of $f_b$ in groups and the field are presented in \S
\ref{sec:results}, and we perform tests with the mock catalogues in
\S~\ref{sec:mock_tests}.  We discuss the implications  of our results
for galaxy 
evolution models in \S \ref{sec:discussion}, and in \S
\ref{sec:conclusion} 
we summarize our results and conclusions.  Readers uninterested in the
details of our selection methods and robustness checks should skip to 
\S\S~\ref{sec:sampledef}, \ref{sec:results} and 
\ref{sec:discussion}.
Throughout the paper we assume a flat, $\Lambda$CDM cosmology with
$\Omega_M=0.3$.

%%%%%%%%%%%%%%%%%%%%%%% SECTION 2 %%%%%%%%%%%%%%%%%%%%%%%%%%%%%

\section{The DEEP2 Galaxy Redshift Survey}
\label{sec:DEEP2}
\subsection{Details of the Survey}

The DEEP2 Galaxy Redshift Survey is the first large, highly accurate
spectroscopic 
survey of galaxies at redshifts around unity.  As of this writing,
the main survey observations are nearly complete ($>95\%$), with spectra
obtained for $49,220$ galaxies in four fields using the
DEIMOS spectrograph on the Keck II telescope.  The survey covers a
total of $\sim 3\; \mathrm{deg}^2$ on the sky to limiting magnitude
$R_{AB}=24.1$.   
The bulk of these observations are of galaxies in the redshift range
$0.7\la z \la 1.4$.  Full details of the survey will appear in the
upcoming paper by Faber et al. (in preparation), but 
summarize the necessary information for this study is summarized
below.

The four fields surveyed were chosen to lie in zones of
low Galactic extinction using the dust maps of \citet{SFD98}.
Three-band 
($BRI$) photometry was obtained for each field 
using the CFH12K camera on the Canada-France-Hawaii Telescope, as
described by \citet{Coil04b}.  Each field is covered by several
contiguous photometric pointings, which are a convenient way to group
and intercompare results.  Three of the DEEP2 fields each cover an
area $90^\prime$ or $130^\prime \times 30^\prime$ on the sky with two
or three contiguous pointings.  In these fields, galaxies are selected
for spectroscopy using a simple cut in $BRI$ colour-colour space that
has been 
optimized to select galaxies at redshifts $z>0.75$.  This cut
efficiently focuses the survey on high-redshift galaxies: it reduces
the portion of the spectroscopic sample at redshifts $z<0.75$ to roughly
$10\%$ while discarding only $\sim 3\%$ of objects at $z>0.75$
(Faber et al. in preparation).  Within each CFHT pointing, galaxies are
selected for spectroscopic observation if it is possible to place them
on one of the $\sim 40$ DEIMOS slit masks covering that pointing.  Slit
masks are tiled in an overlapping chevron pattern using an adaptive
algorithm 
to increase the coverage in dense regions on the sky, giving nearly
every galaxy two chances to be placed on a mask.  Further details of
the observing scheme are given in \citet{DGN04}; overall DEEP2 targets
$\sim 60\%$ of galaxies that meet its selection criteria.

The fourth field of the survey, the
Extended Groth Strip (hereafter EGS), covers $120^\prime \times
16^\prime$ on the sky.  A concerted effort is underway by a large
consortium of observing teams (the AEGIS team; \citealt{AEGIS}) to
obtain a wide array of 
observations of this field from X-ray to radio wavelengths.
Therefore, to maximize the evolutionary 
information that will be available, galaxies in the EGS
have been targeted for spectroscopy regardless of estimated redshift,
and at a 
significantly higher sampling rate than in other fields: each galaxy
in EGS has four chances to be
selected for observation rather than two.  This means that galaxies
selected for 
spectroscopy in the EGS constitute a superset of galaxies that would
have been selected using the criteria of the other three DEEP2 fields;
hence it is possible to create a high-redshift subsample of the EGS whose
selection (including sampling rate) is identical to that of the rest
of the DEEP2 survey.  We will 
include this subsample in the present study.  

At this writing, spectroscopic observations have been completed for all
three of the high-redshift DEEP2 fields
and for more than three-quarters of the EGS field.
All DEEP2 spectra have been reduced using an automated data-reduction
pipeline (Cooper et al. in preparation), and redshift identifications
are all confirmed visually.   Rest-frame $U-B$ colours and absolute $B$
magnitudes are 
computed using the K-correction algorithm described in Appendix A of
\citet{Willmer06}.  This study uses data from all of
the DEEP2 CFHT pointings for which spectroscopic observations have
been completed.
In each pointing, the fraction of the spectra that yield a
successful redshift is greater than $70\%$.

\subsection{The Group Catalogue}
\label{sec:groupcat}

The details of the DEEP2 group-finding procedure are fully discussed
in \citet{Gerke05}; we have now applied this procedure to the full
current DEEP2 dataset, which is significantly larger than that used in
the previous paper.   We summarize the salient points of the
group-finding algorithm here.  

Groups of galaxies in the DEEP2 sample are identified using the
Voronoi-Delaunay Method (VDM) group finder, which was originally
implemented by \citet{MDNC02}.  This group finder searches adaptively
for groups (bound, virialized associations of two or more observed 
galaxies) using information about local density derived from 
the Voronoi partition and Delaunay complex of a given
three-dimensional galaxy sample.  \citet{Gerke05} calibrated the VDM 
group finder using the mock catalogues of \citet{YWC03}, achieving the
primary
goal of accurately reconstructing the bivariate distribution
$n(\sigma, z)$ 
of groups as a function of redshift and velocity dispersion for 
dispersions $\sigma \ge 350 \kms$.  

For the purposes of this paper, 
however, we will be considering the properties of galaxies, rather
than group properties,  so we will focus
on somewhat different measures  
of success here.  In particular, this paper studies properties of
the population of galaxies within groups (the \emph{group sample})
and of the population of isolated galaxies (the \emph{field sample}),
so it is crucial to determine the success of the VDM at identifying
each of these populations.  
By testing the VDM group finder on mock DEEP2 catalogues, \citet{Gerke05}
showed that  
the fraction of real group members that are successfully identified 
as such (the \emph{galaxy success rate}, $S_{\mathrm{gal}}$) is 
$0.79$.  Conversely, the fraction of galaxies in the
reconstructed group population that are actually misclassified field
galaxies (the \emph{interloper fraction}, $f_I$) is $0.46$.  In
addition, $82\%$ of field galaxies are correctly identified, while
only $6\%$ of the reconstructed field  sample is made up of 
misclassified group members.  
Both samples (group and field) are thus dominated by correctly
classified galaxies, but each sample is contaminated by galaxies from
the other.  Therefore, any differences between the group and field
galaxies should be somewhat stronger in reality
than what the VDM reconstructs.   As will be discussed below in
\S~\ref{sec:sampledef}, the contamination of the group sample is
particularly bad for groups with velocity dispersion below $100 \kms$,
so we will reclassify galaxies in these groups as field galaxies.

Finally, it is important to note here that groups in the DEEP2 survey
are typically of modest mass.
The mock catalogues of \citet{YWC03} indicate that 
the bulk of the DEEP2 groups should have virial masses in the range
$5\times 
10^{12} \la M_\mathrm{vir} \la 5\times 10^{13} M_\odot$ ($200 \la
\sigma_v \la 400 \kms$), with very few groups having $M_\mathrm{vir} >
10^{14} M_\odot$; this is consistent with the estimates of the minimum
group mass derived from the autocorrelation function of DEEP2 groups
\citep{Coil06}. In what follows, then, any conclusions drawn about
the properties of groups should not be taken to apply to rich
clusters. 

\section{Mock Catalogues}
\label{sec:mocks}
The study of groups and clusters of galaxies is fraught with
unavoidable sources of systematic error.  For example, it
has been shown \citep{SzSz96} that an error-free cluster catalogue is
not achievable for an incompletely sampled galaxy distribution, even
if the physical distances to the galaxies are known.
Moreover, in a redshift survey the peculiar velocities of galaxies
induce distortions in the redshift-space distribution of
galaxies, mixing the positions of galaxies in 
groups with the positions of isolated galaxies and  making accurate
group 
detection even more difficult.  Because of these unavoidable sources
of systematic error, robust conclusions require
that we test our methods on realistic mock galaxy catalogues.

\citet{Gerke05} tested and calibrated their group-finding
methods with the mock 
catalogues of \citet{YWC03}.  These catalogues were produced by populating
dark-matter-only N-body simulations with galaxies following a halo
model prescription that follows \citet{YMvdB}.  In
particular, the mocks are populated using 
a conditional luminosity function, $\phi(L|M)$, that assigns galaxies to
a halo of mass $M$ according to a luminosity function whose parameters
depend on $M$.  The parameters of the model were chosen to
be consistent with the two-point correlation function $\xi(r)$
observed in early DEEP2 data \citep{Coil04} and with the local
$\xi(r)$ from Peacock et al.~\citep{Peacock01}.  These mocks were
sufficient for the purpose of testing our overall success at
reconstructing groups, but they do not
provide information about  
galaxy properties aside from luminosity.  

In the current work we shall need to test 
our success at reproducing the colours of the
group galaxy population; this requires mock
catalogues that include that information.  
The mocks must also reproduce any dependences 
of the galaxy colour distribution on local overdensity, since that is
the trend
we aim to probe in the data.  To this end, we assign colours to
mock galaxies by drawing colours from galaxies in similar environments
within the 
actual DEEP2 data.  This method is similar to, but less sophisticated
than, the one used by Wechsler et al. (in preparation) to create the
mock SDSS catalogues that were used to test the C4 group-finding algorithm
\citep{Miller05}.  

%\subsection{Adding colour information to the mocks}
The
procedure is as follows.  Local galaxy density in the
DEEP2 data is measured by
computing the distance (projected on the sky) to each galaxy's
third-nearest neighbor within $1000\kms$ in redshift, as detailed in   
\citet{Cooper05}.  DEEP2 galaxies contaminated by
edge effects are discarded, and the data sample is limited to galaxies
in the range $0.8<z<1.0$ 
to minimize the effect of the apparent magnitude limit.  In the mock
catalogue, local density is estimated using the projected distance to the 
\emph{seventh}-nearest neighbor (within a catalogue of an enhanced spatial
extent, to avoid edge effects in the final catalogue). The
seventh-nearest neighbor in the complete, volume-limited mock
catalogue has been shown to be a reasonable analogue to the third-nearest
neighbor measured 
in the more sparsely sampled magnitude-limited data
\citep{Cooper05}. We then divide 
both the mock and data samples into quintiles of local density and
identify each quintile in the data  
with the quintile in the mock that has the same local-density rank.
The colour distributions within each density bin in the
data can then be used to assign colours to the mock galaxies. For
example, the $20\%$ of 
mock galaxies with the highest local densities will be assigned colours
randomly drawn from the observed colour distribution of the
highest-density 
$20\%$ of DEEP2 galaxies.  

Before doing this, however, we also divide the samples by luminosity,
since DEEP2 galaxy colour is observed to be correlated with
luminosity (see, e.g., \citealt{Willmer06}).   For
the DEEP2 data, we sort galaxies into four bins in absolute $B$ band
magnitude.  To ensure that similar parts of the 
luminosity function are being considered at all $z$, we shift these
bins with redshift to 
account for evolution in the typical galaxy magnitude $M^\ast$.
\citep{Faber06} found that this value evolves as $M^\ast\propto Qz$,
with $Q=-1.37$; we apply the same linear function of $z$ to our
luminosity binning of the data.     
The bins include only galaxies brighter than
$M_B-5\log{h}+Q(z-1)=-20$, 
since the sample is incomplete for fainter magnitudes (see
Figure~\ref{fig:colmag}), and each bin is 0.5 magnitudes wide, except
for the brightest bin, which includes all galaxies brighter than
$M_B-5\log{h}+Q(z-1)=-21.5$.  The mock catalogue is divided into the
same four bins, 
except that in this case the bins evolve with redshift as $M\propto
Q_\mathrm{mock}z$, with $Q_\mathrm{mock}=-1$, the $M^\ast$
evolution parameter that was 
assumed in \citet{YWC03}.  Also the faintest bin in the mocks includes
all galaxies fainter than $M_B+Q_\mathrm{mock}(z-1)=-20.5$, which
means that all mock galaxies fainter 
than this limit will have the same colour distribution as DEEP2 galaxies
in the range $-20.5<M_B-Q(z-1)<-20$.  This is the best that can be
done using this procedure, since the DEEP2 sample is  
incomplete for fainter objects; in any event, most of the mock galaxies
in this regime will fall below the DEEP2 apparent magnitude limit, so
there is little practical effect.

\begin{figure}
\begin{center}
\epsfig{width=0.77\linewidth, angle=90, file=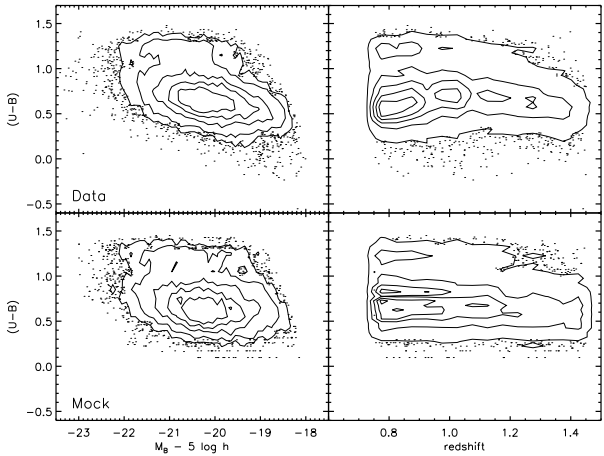}
\caption{Colour-magnitude and colour-redshift diagrams for the DEEP2
  survey and for the mock catalogues described in the text.  To aid
  comparison, the mock  
  sample has been randomly diluted to contain the same number of
  galaxies as the data set, and each of the two data plots has the
  same contour levels as the corresponding mock plots (contours are
  spaced evenly in density).
  The subtle correlation between colour and luminosity in the data is
  reproduced well in the mocks at $M_B-5\log{h}<-20$; at fainter
  magnitudes it 
  is absent by construction.}
\label{fig:chromo_cmd}
\end{center}
\end{figure}

Having divided the samples thus into bins of
local overdensity and absolute magnitude, we then add colours to the mock
population by considering galaxies in each bin separately.  For each
mock galaxy in a given luminosity-density bin, a
real galaxy is selected at random from the 
corresponding bin in the DEEP2 data sample, and we assign that
galaxy's rest-frame $U-B$ colour to the mock galaxy in question.  This 
procedure produces a distribution in $M_B$ vs. $U-B$ colour-magnitude
space that matches the distribution observed in DEEP2 reasonably well
(see Figure~\ref{fig:chromo_cmd}). Because the procedure
only uses DEEP2 galaxies in the range $0.8<z<1.0$ and because it does
not divide the samples by redshift, any intrinsic evolution in the
colours of DEEP2 galaxies will not be reproduced in the mock catalogues.
This is desirable, however, since it will allow us to confirm that our
selection and group-finding procedures have not introduced any
spurious evolutionary trends.   

Using the colours assigned to the mock galaxies, it is possible to the
K-correction procedure described in 
\citet{Willmer06} to assign each galaxy an $R$-band apparent
magnitude, which can be used to select mock galaxies with the same
$R<24.1$ apparent magnitude limit that is used in DEEP2.
In addition to this magnitude limit, we also apply the DEEP2
DEIMOS slitmask-making algorithm to the mock catalogue, as projected on
the sky, removing those galaxies that would not
be targeted for observation.  Finally, we dilute the remaining galaxy
sample to reproduce the $\sim 70\%$ redshift success rate of the
survey; the dilution procedure accounts for the slight magnitude
dependence of this rate.

%%%%%%%%%%%%%%%%%%%%%%%%%%%%%%%% SECTION 3 %%%%%%%%%%%%%%%%%%%%%%

\section{Sample definition and measurement methods}
\label{sec:selection}
\subsection{Defining the galaxy sample}
\label{sec:sampledef}

Because the DEEP2 survey extends over a broad redshift range ($\Delta
z \sim 0.7$), selecting galaxies according to a simple apparent
magnitude limit introduces selection effects that depend strongly on a
galaxy's rest-frame colour.  For example, the DEEP2 $R_{AB}\le 24.1$
magnitude limit corresponds to selection in the rest-frame $B$ band at
$z\sim 0.7$ and the rest-frame $U$ band at $z\sim 1.2$.  Therefore,
galaxies with intrinsically red colours will fall beyond the selection
cut at lower redshifts than those with intrinsically blue colours.

\begin{figure*}  
\begin{center}
\epsfig{width=0.8\linewidth, file=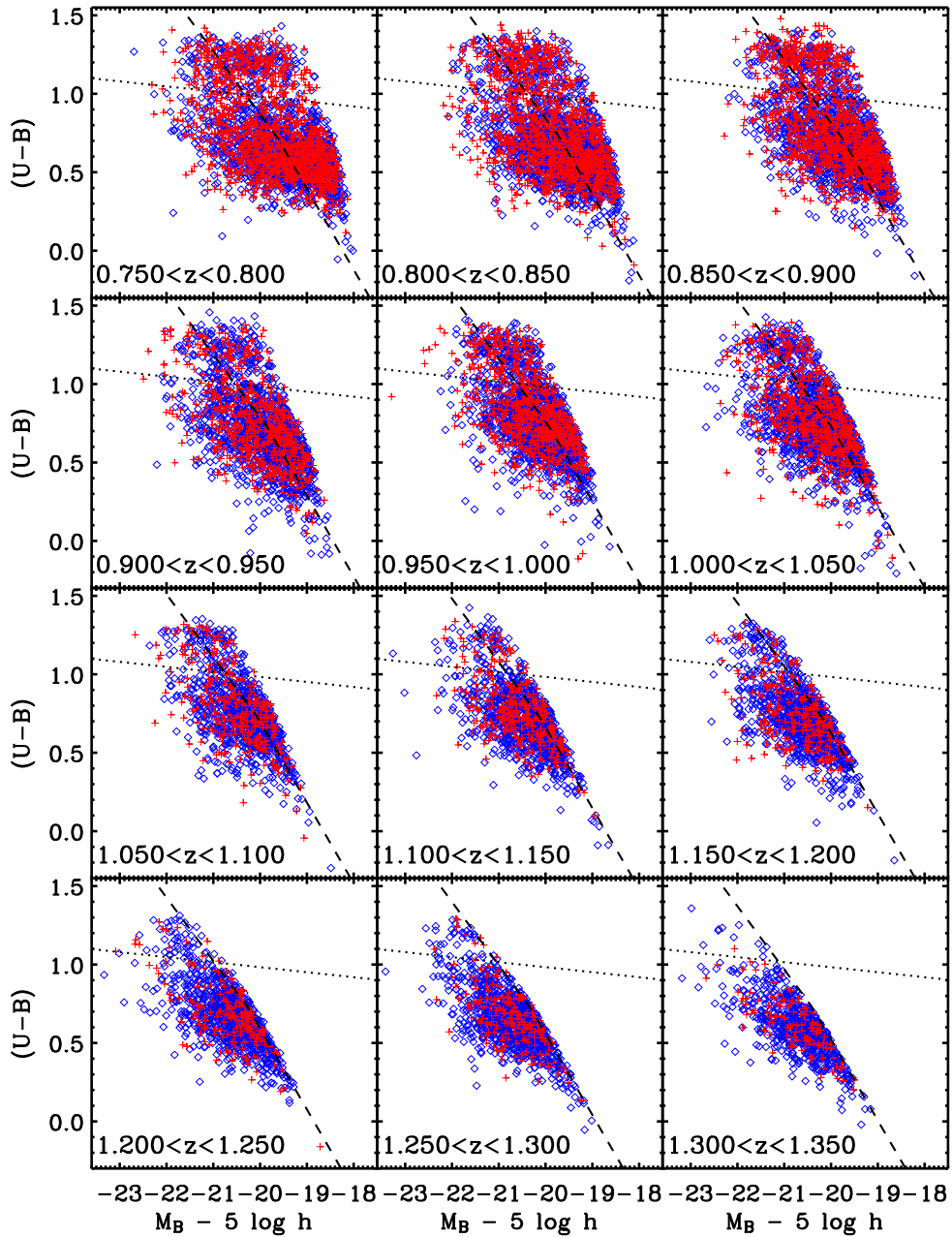}
\caption{Rest-frame colour-magnitude diagrams for DEEP2 galaxies in
redshift bins 
of width $\delta z=0.05$.  Crosses (red) denote galaxies in groups, and
diamonds (blue) show field galaxies. Groups that have fewer than two
members above the magnitude limit have not been excluded from the
group sample in this plot, although they are excluded elsewhere in the
paper, as discussed in the text.  The dashed lines show the effective 
DEEP2 apparent magnitude limit, as given in Equation~\ref{eqn:magcut},
using parameters from Table~\ref{tab:samples} for $z_{\mathrm{lim}}=1.3$.  Dotted
lines indicate the division between blue and red galaxies as given in
Equation~\ref{eqn:colourcut}.}
\label{fig:colmag}
\end{center}
\end{figure*}

This effect is readily apparent in Figure~\ref{fig:colmag}, which shows
rest-frame colour-magnitude diagrams ($M_B$ vs. $(U-B)$) for DEEP2
galaxies, divided into 
redshift bins of width $\Delta z = 0.05$.  In each panel, distinct red
and blue galaxy populations are apparent, with loci that are roughly
divided by the dotted lines.  The sharp cutoff in the
galaxy population on the right side of each panel is caused by the
DEEP2 apparent magnitude limit; this selection cut becomes
increasingly biased against red galaxies as redshift increases and the
observed $R$ band moves further blueward of the rest-frame $B$ band. 

It is obviously necessary to take this effect into account when
studying the redshift evolution of the blue fraction.  In particular,
we must ensure that galaxies of a given colour have been equally well
sampled at all redshifts being considered. 
The simplest selection method is to produce a volume-limited catalogue
with an absolute magnitude limit---\emph{i.e.}, a vertical selection
cut in colour-magnitude space.
For DEEP2, however, this
method severely restricts either the redshift range probed or the
number of galaxies selected.  For example, as can be seen in
Figure~\ref{fig:colmag}, a limiting absolute magnitude of $M_B - 5\log{h}=-20$ 
would allow measurement of $f_b$ only out to redshift $z\sim 0.9$; beyond
this, the red galaxy sample would be incomplete, resulting in a
spurious sharp rise in $f_b$ at all higher redshifts.  On the other
hand, a limiting magnitude of $M_B-5\log{h} = -21.7$ would give a
complete, 
volume-limited sample out to $z=1.3$, but such a sample would contain
far too few galaxies for a robust measurement of $f_b$.

However, to examine the evolution of galaxy properties, 
all that is required is to select a region of
colour-magnitude space that is uniformly sampled by the survey at all
redshifts of interest.
Such a selection cut is shown by the dashed curves in each panel of 
Figure~\ref{fig:colmag}, which are described by the equation
\begin{eqnarray}
\label{eqn:magcut}
M_{\mathrm{cut}} -  5\log{h}&=&  
  Q(z-z_{\mathrm{lim}}) \\
     &+& \mathrm{min}\left\{[a(U-B) + b],\ [c(U-B) +
        d]\right\}, \nonumber
\end{eqnarray}
where $z_{\mathrm{lim}}$ (equal to $1.3$ in the figure) is the
limiting redshift beyond
which the selected sample becomes incomplete, $a, b, c$, and $d$ are
constants that depend on $z_{\mathrm{lim}}$ and are determined by
inspection of the colour-magnitude diagrams, and $Q$
is a constant that allows for 
linear redshift evolution of the typical galaxy luminosity $L^\ast$,
as already mentioned in \S~\ref{sec:mocks}. 
\citet{Faber06} have measured the evolution of the galaxy luminosity
function using data from COMBO-17~\citep{Bell04b}
and DEEP2 at $z\sim 1$ in conjunction with low-$z$ data from the
2dFGRS \citep{Madgwick02, Norberg02} and SDSS \citep{Bell03,
  Blanton03}; they 
find that changes in $L^\ast$ are   
well described by a linear evolution model, $L^\ast\propto Qz$, with
$Q=-1.37$.  By  
including this evolution in our selection cut, we are selecting a
similar population of galaxies  \emph{with respect to
$L^\ast$} at all redshifts.

\begin{figure}
\centering
\epsfig{width=\linewidth, file=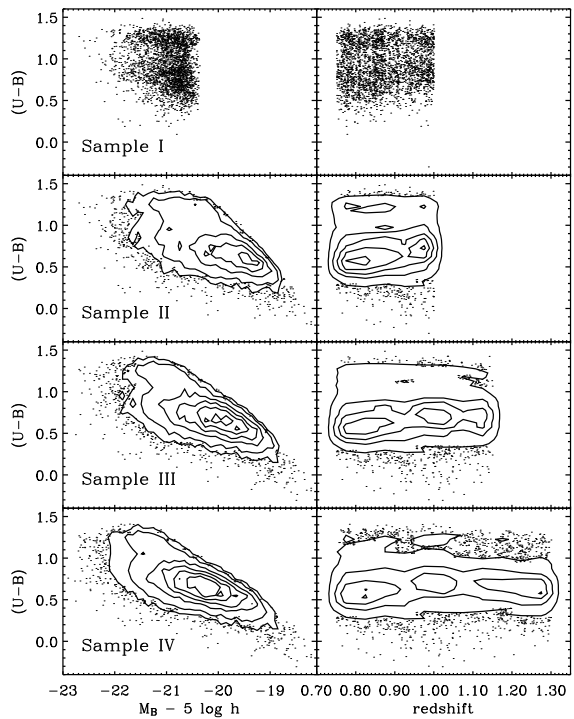}
\caption{Summary of the galaxy catalogue in each of the four samples
  defined in Table~\ref{tab:samples}.  The left-hand panels show
  colour-magnitude diagrams for the samples, and the right-hand panels
  show colour-redshift plots.  The redshift limits of the samples are
  apparent, as are 
  the effects of our colour-magnitude selection criteria
  (Equation~\ref{eqn:magcut}).  In the bottom three rows, contour lines are
  evenly spaced in point density.}
\label{fig:samplegals}
\end{figure}

We use Equation~\ref{eqn:magcut} to define three different samples,
called samples II, III and IV, which are
complete in colour-magnitude space to $z=1.0, 1.15$ and $1.3$,
respectively; the parameters 
defining these samples are given in Table~\ref{tab:samples}.  In
addition, we create a sample, called sample I, that is purely
absolute-magnitude-limited 
(relative to $M^\ast$) and complete to $z=1.0$ by applying a simple
cut at $M_B - 5\log{h}=-20.7+ Q(z-1.0)$. 
Table~\ref{tab:samples} summarizes the properties of the resulting
samples, and Figure~\ref{fig:samplegals} shows
colour-magnitude and colour-redshift diagrams for the galaxies in each
sample.  These samples should also be useful
for future studies of colour evolution in DEEP2.

Selecting galaxy samples according to Equation~\ref{eqn:magcut},
effectively creates a sample that is volume-limited \emph{for each
colour}.  That is, the selection is colour-dependent, but the magnitude
limit relative to $M^\ast$ is redshift-independent.  This selection is
different than the traditional 
volume-limited selection, but it nevertheless allows for comparison of
the relative numbers of galaxies of different colours at different
redshifts; that is, it allows us to examine the evolution of the blue
fraction.  It should be noted, however, that our values of $f_b$
cannot be meaningfully compared to published values at low redshift,
which have typically been computed using a simple magnitude cut.  In
particular, our values of $f_b$ will be much higher than local values
from the literature since the magnitude cut in Equation~\ref{eqn:magcut}
selects many more blue galaxies than red.  Indeed, $f_b$ values cannot
even be compared among the different samples defined in
Table~\ref{tab:samples}.  The absolute values of $f_b$ in each sample
are mainly set by the selection; useful information comes only
from splitting these samples into subsamples to look for \emph{trends} with
redshift or with galaxy properties.

Before we move on, it is important to note that the application
of our colour-magnitude selection criteria has important implications
for the 
definitions of our group and field samples.  The VDM group catalogue has
been defined using the entire apparent-magnitude-limited survey.  It
is thus possible that a group could have been detected with two or
more members at $z=0.8$, say, while only one of these members is brigher
than $M_\mathrm{cut}$.  Had this group instead been at $z=1.3$, it
clearly would not have been identified, and its one bright galaxy
would have been considered to lie in the field.  If no correction is
made for this effect, the 
application of $M_\mathrm{cut}$ will cause a redshift nonuniformity in
our samples, with galaxies from relatively faint groups being
counted as group members at low $z$ and as field members at high $z$.
To avoid such biases, we redefine the group sample to be only those
galaxies that reside in groups with two or more members brighter than
$M_\mathrm{cut}$.  The field sample comprises all other
galaxies.   

\begin{figure}
\centering
\epsfig{width=\linewidth, file=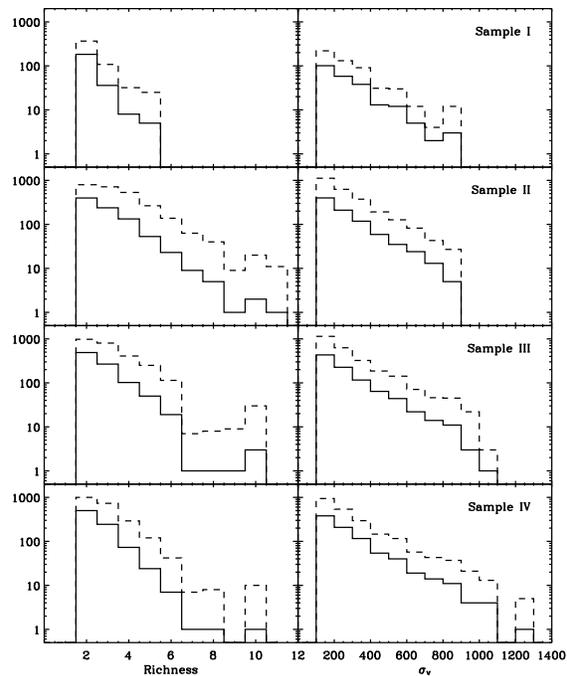}
\caption{Summary of the group catalogues in each of the four
  samples defined in Table~\ref{tab:samples}.  Solid lines show the
  differential distribution of groups as a function of richness
  (number of galaxies above $M_\mathrm{cut}$ for the sample) and
  velocity dispersion $\sigma_v$.  Dashed lines show the total numbers of
  \emph{galaxies} within those groups that fall in a given bin. }
\label{fig:samplehists}
\end{figure}

We also define a group's richness $N$ to be the number of
galaxies with $M_B \le M_\mathrm{cut}$ instead of the total number of
observed galaxies, so that richness is measured consistently at all
$z$ (other group properties, like $\sigma_v$ and mean $z$,
are still computed using \emph{all} observed group members, for the
sake of robustness). 
This group definition ensures that our sample of group
galaxies is drawn from comparable groups at all $z$, although it means
that a given group may have a different value of $N$ in each
sample. Also, throughout 
this paper we will restrict the group galaxy sample to those galaxies
whose host groups have velocity dispersions $\sigma_v\ge 100 \kms$.
This is because, as shown in \citet{Gerke05} (see Figure 8 of that
paper), groups with lower velocity dispersion are predominantly false
detections.  We therefore class galaxies in such groups with the
field sample.  This point is discussed further in
\S~\ref{sec:fb_local_trends}.  Figure~\ref{fig:samplehists} shows the
distribution of groups and their member galaxies as a function of $N$
and $\sigma_v$ for each of the four samples.

\begin{table*}
  \centering
  \begin{minipage}{140mm}
    \caption{Summary of the data samples}
    \begin{tabular}{@{}lcccc@{}}
%      \tabletypesize{\scriptsize}
%      \tablewidth{0pt}
%      \tablecaption{Summary of the data samples}
\hline
\hline
%      \tablehead{
%\colhead{} &\colhead{sample I} & \colhead{sample II}& \colhead{sample III}  &
%\colhead{sample IV}
%}
     & Sample I & Sample II & Sample III & Sample IV\\ 
%\startdata
\hline
Description  &  Simple $M_B-M^\ast$  &  Colour-magnitude & Colour-magnitude &
Colour-magnitude \\
             &  limit to z=1.0 &  complete to z=1.0 & complete to z=1.15 & complete
             to z=1.3 \\
%\tableline
\hline
$z_{\mathrm{lim}}$\footnote{The parameters $z_{\mathrm{lim}}, a, b, c$, and $d$ are defined in
  Equation~\ref{eqn:magcut}.}  & 1.0    & 1.0    & 1.15   & 1.3  \\
$a$ & 0      &  -1.34 &  -1.55 & -1.94 \\
$b$ & -20.70 & -18.55 & -18.77 & -18.92 \\
$c$ &  ...   &  -2.08 &  -2.32 & -2.90 \\
$d$ &  ...   & -17.75 & -18.16 & -18.53 \\
\hline
\# galaxies         &   2691 &   9546 &  11767 &  12493 \\
\# groups\footnote{Groups must have two or more members above the
  sample's magnitude limit.}          &    232 &    863 &    933 &    851 \\
\# group galaxies%\tablenotemark{2}  
                  &    531 &   2588 &   2605 &   2211 \\
%\tableline
\hline
overall $f_b$\footnote{See Equation~\ref{eqn:f_b} for the definition of the
  blue fraction, $f_b$.}    & 0.603 $\pm$ 0.011 & 0.827 $\pm$ 0.005 & 0.877 $\pm$ 0.004 & 0.924 $\pm$ 0.003 \\
field $f_b$       & 0.624 $\pm$ 0.012 & 0.854 $\pm$ 0.005 & 0.893 $\pm$ 0.004 & 0.933 $\pm$ 0.003 \\
group $f_b$        & 0.517 $\pm$ 0.025 & 0.749 $\pm$ 0.010 & 0.818 $\pm$ 0.008 & 0.876 $\pm$ 0.008 \\
%\enddata
\hline
%\tablenotetext{1}{The parameters $a, b, c$, and $d$ are defined in
%  Equation~\ref{eqn:magcut}.}
%\tablenotetext{2}{Groups must have two or more members above the
%  sample's magnitude limit.}
%\tablenotetext{3}{See Equation~\ref{eqn:f_b} for the definition of the
%  blue fraction $f_b$.}
\label{tab:samples}
%\end{deluxetable*}
\end{tabular}
\end{minipage}
\end{table*}

\subsection{Correcting for incompleteness}
\label{sec:weights}

Before we may proceed with measuring $f_b$, an
additional effect must be accounted for.  Because of the finite
slit length available on the DEIMOS spectrograph, DEEP2 can
target for spectroscopy only $\sim 60\%$ of the galaxies that meet its
selection criteria
(we will call the remaining potential targets ``unobserved galaxies'').
Moreover, $\sim 30\%$ of  
galaxies that are targeted for spectroscopic observation fail to yield
redshifts (we will call these ``redshift failures'').  Follow-up
observations have shown that many ($\sim 50\%$)  of 
the redshift failures, especially for blue galaxies, lie at redshifts
beyond the range probed by 
DEEP2 (C. Steidel, private communication).  The overall failure rate is
also correlated with observed galaxy colour and magnitude, although the
trends are only slight.   Other failures may occur
because of poor observing conditions, data reduction errors (e.g., poor
subtraction of night-sky emission), or
instrumental effects (e.g., bad CCD pixels), but such problems
are only a minor contribution to the overall failure
rate. These various sources of incompleteness  may introduce spurious 
evolutionary trends in $f_b$ unless an appropriate correction is applied.

In this work we
adopt the corrective weighting scheme used by
\citet{Willmer06} in measuring the DEEP2 luminosity function.  A method
of this sort was first implemented by \citet{Lin99} for the CNOC2
Redshift survey. We locate each galaxy in the three-dimensional space
defined by observed $R$ 
magnitude and  $R-I$ and $B-R$ colours.  We then define a cubical bin
(0.25 magnitudes on a side)
in this space around each galaxy with a successful
redshift and compute a weight for the $i$th such galaxy:
\begin{equation}
\chi_i= 1 + \frac{\sum_j P_i(z_{\mathrm{lo}}\leq z_j \leq
                  z_{\mathrm{hi}})}{N_z^i}.
\label{eqn:weight}
\end{equation} 
Here $N_z^i$ is the number of galaxies in the bin around galaxy $i$
with successful 
redshifts within the nominal DEEP2 redshift 
range ($z_{\mathrm{lo}}=0.7\leq z\leq 1.4=z_{\mathrm{hi}}$), $j$ is an
index that runs over all galaxies in the bin for which DEEP2 did not
successfully measure a redshift (whether they were observed or not),
and $P_i(z_{\mathrm{lo}}\leq z_j \leq z_{\mathrm{hi}})$ 
denotes the probability that galaxy $j$ (in the bin around galaxy $i$)
falls in the nominal DEEP2 redshift range.  

To compute the probability $P$ in equation~\ref{eqn:weight}, we must
construct a model for the redshift distributions of redshift failures
\emph{and} of unobserved galaxies.  
Modeling this in detail would require additional assumptions, but it
is reasonably certain that the 
truth lies between two extreme models.  As noted previously, many of
the failures are known to be at $z>1.4$, so we can start by making
the extreme assumption that \emph{all} failures lie in this range.  In
this case, called the ``minimal'' model, all redshift failures have
$P_i(z_{\mathrm{lo}}\leq z_j \leq  z_{\mathrm{hi}}) =0$ in
equation~\ref{eqn:weight}, and unobserved galaxies have 
\begin{equation}
P_i(z_{\mathrm{lo}}\leq z_j \leq  z_{\mathrm{hi}}) = 
        \frac{N^i_z}{N^i_z+N^i_{zl}+N^i_{zh}+N^i_f},
\end{equation}
where $N^i_{zl}$ is the number of galaxies (in the $i$th bin in
colour-colour-magnitude space) with redshifts observed to be
below $z_{\mathrm{lo}}$,  $N^i_{zh}$ is the number with
$z>z_{\mathrm{hi}}$, and $N^i_f$ is the number of observed galaxies
that failed to yield a redshift.  Taking the opposite extreme, we
could assume that redshift failures have exactly the same redshift
range as the galaxies with successful redshifts.  In this case, called
the ``average'' model, both redshift failures and unobserved galaxies
have the same value for $P_i$: 
\begin{equation}
P_i(z_{\mathrm{lo}}\leq z_j \leq  z_{\mathrm{hi}}) = 
        \frac{N^i_z}{N^i_z+N^i_{zl}+N^i_{zh}}.
\end{equation} 
Since blue galaxies with failed redshifts are known to be
frequently beyond the DEEP2 redshift range, \citet{Willmer06} adopted
a compromise model (the ``optimal'' 
model), in which blue galaxies (see \S~\ref{sec:computing_fb} for
the precise definition) are corrected with the minimal model 
while red galaxies are weighted using the average model.  We shall
adopt this scheme in what follows; however, our basic conclusions are
insensitive to the weighting scheme used and in fact are unchanged
even when no weighting is used.

\subsection{Computing the blue fraction}
\label{sec:computing_fb}
Having defined a galaxy sample using equation~\ref{eqn:magcut} and a
set of galaxy weights $\chi_i$ using equation~\ref{eqn:weight}, we may
now compute $f_b$ for our chosen sample.  As in \citet{Willmer06} we
divide the galaxies into 
red and blue subsamples according to the observed bimodality in galaxy
colours.  This division corresponds to the dotted line shown in all
panels of Figure~\ref{fig:colmag}, given by the equation
\begin{equation}
U-B= -0.032(M_B-5\log{h}+ 21.62) + 1.285 - 0.25,
\label{eqn:colourcut}
\end{equation}
which was derived from the \citet{vanD00} colour-magnitude relation for
red galaxies in distant clusters.  Eqn.~\ref{eqn:colourcut} shifts that
relation downward by 0.25 magnitudes to pass through the valley in the
colour distribution, following
\citet{Willmer06}. One can
also allow this colour division to evolve with redshift according to
passive stellar evolution models; however, there is no evidence in
the DEEP2 data for any
evolution in the position of the valley (See Figure~\ref{fig:colmag}),
and a realistic amount of 
evolution has only minimal effects on our results (see \S 5.2).  In
the interest of  
simplicity, therefore, we do not allow the division to evolve in most
of what follows.
This division between red and blue galaxies produces a set of
$M$ blue galaxies out of a total sample of $N$ galaxies, each of which
is assigned a weight $\chi_i$.  The corrected blue fraction is then
given by 
\begin{equation}
f_b=\mN_b/\mN_t,
\label{eqn:f_b}
\end{equation}
where the number of blue galaxies $\mN_b$ and the total
number of galaxies $\mN_t$ are defined as 
\begin{equation}
\mN_b=\sum_{j=1}^M \chi_j,\; \mN_t=\sum_{i=1}^N\chi_i
\label{eqn:N_def}
\end{equation}
with the index $j$ running over all of 
the blue galaxies and the index $i$ running over the  
full galaxy sample.  

\subsection{Estimating errors on $f_b$}
\label{sec:fb_errors}

The formal error in $f_b$ is given by simple binomial statistics: 
\begin{equation}
\sigma^2_\mathrm{bin}(f_b) = \frac{\mN_b(\mN_t-\mN_b)}{\mN_t^3}
\end{equation}
(except in the case $\mN_b=0$, for which
\citet{DeP04} argue that $\sigma(f_b) = 1/(2\mN_t)$).  However, this
formula will not fully account for the scatter in our measured values
of $f_b$.  Because a correlation exists in DEEP2 between local galaxy
density and galaxy colour \citep{Cooper06a}, large-scale
structure will induce an intrinsic scatter in values of $f_b$ measured
over a finite volume, in addition to
the formal binomial error; this can be thought
of, in essence, as a contribution to the error from cosmic variance.
We discuss the effects of this scatter further in
\S~\ref{sec:fb_evol}.  Also, errors  
in the galaxy weights $\chi_i$ will add scatter to the measured $f_b$
values.  We therefore find it most convenient to estimate our errors
empirically.  

We do this in three different ways to ensure that our methods are
robust.  First, we compute $f_b$ for each of the 10 DEEP2 photometric
pointings 
individually and average those computed values; the error
$\sigma(f_b)$ is then taken to be the standard error on that mean.
Second, we estimate $\sigma(f_b)$ using a jackknife sampling strategy
in which each pointing is removed in turn from the full dataset being
considered, $f_b$ 
is computed for each subsample, and the error on $f_b$ for the full
sample is estimated using the usual jackknife error formula.
Finally, we compute the errors on $f_b$ using a bootstrap resampling
strategy in which the entire data subsample under consideration
(\emph{e.g.} all group galaxies in a given redshift bin) is randomly  
sampled with replacement, and the error on $f_b$ is estimated from the
standard deviation of the $f_b$ distribution for 500 such Monte-Carlo
samples, using standard 
bootstrap methods.  In each case, the estimated error must be scaled
up by a 
factor of $1.08$ to account for the covariance that 
exists between contiguous pointings due to large-scale structure
fluctuations (\emph{i.e.}, cosmic variance).  This factor is derived from
Monte Carlo tests based on the covariance between fields with the
actual DEEP2 geometry, using the cosmic variance calculation code of
\citet{ND02}; it corresponds to the conservative assumption that
pointing-to-pointing variations in $f_b$ are dominated by cosmic
variance. 

These three different methods give comparable results, though the
standard-error and jackknife estimates are much noisier than those
from bootstrapping.  In the interest of stability, we will 
always report the bootstrap error values in what follows.  It is,
however, worth emphasizing that even this method may not account for
all sources of 
$f_b$ variance in a given redshift bin, since the DEEP2 dataset is
finite.   Because $f_b$ seems to be lower in higher-density regions, if
we are particularly unlucky and there is a net overdensity or 
underdensity at a given $z$ in most of the DEEP2 fields, this may lead
to a 
fluctuation in $f_b$ that is not accounted for in our error bars.  In
particular, it appears that there may be an underdensity at $z\sim
0.9$ in most of the DEEP2 fields; this may lead to unusually high
values of $f_b$ at that redshift.  To ameliorate this problem, in
addition to computing $f_b$ in independent bins, we will also
sometimes compute $f_b$ in a sliding box, which will smooth out any
large-scale structure fluctuations at the expense of introducing
bin-to-bin correlations in the resulting measurements.

\section{Results}
\label{sec:results}
Table~\ref{tab:samples} shows the properties of the four samples 
defined in \S~\ref{sec:sampledef}, including blue fractions.  
A primary result of this work is apparent in the last two rows of 
the Table: in all four samples considered, the blue fraction is
significantly lower in groups than it is in the field.
Thus, the well-known \emph{qualitative} distinction between local 
field and group (or cluster) galaxy populations was in place by $z\sim
1$.   
%Since the main stated purpose of this work is to study the
%quantitative evolution of $f_b$, however, 
It is, however, worth emphasizing  that the \emph{quantitative}
increase in $f_b$ values from sample I to sample IV is \emph{not}
evidence for evolution because the
overall value of $f_b$ in each sample is strongly affected by the
sample-selection procedures in \S~\ref{sec:sampledef} (\emph{i.e.},
sample IV has a much more sharply tilted selection cut than sample II,
so it will include proportionally fewer red galaxies by construction).
In what follows, we will divide each 
sample into redshift bins, allowing for a meaningful investigation of
evolutionary trends in the blue fraction.  First,
though, it is important to understand the trends that
exist at \emph{fixed} redshift. 

\subsection{Blue-fraction trends at fixed redshift}
\label{sec:fb_local_trends}

As discussed in the introduction, the blue fraction in local clusters
exhibits a large system-to-system scatter that arises,
at least in part, because the measured value of $f_b$
depends on exactly \emph{which} galaxies and clusters are used to
compute it (e.g., \citealt{DeP04, Poggianti06}). 
Before considering $f_b$ evolution
in DEEP2, then, it is crucial to understand trends in the blue
fraction at \emph{fixed} redshift; otherwise they may introduce
selection effects that enhance or  
compete with evolutionary effects (\emph{e.g.},
\citealt{Smail98}).   
As discussed in
\S~\ref{sec:fb_evol} below, there is little to no evolution in $f_b$
over the redshift range $0.75<z<1.0$ in DEEP2, so we will limit
ourselves to this range (\emph{i.e.} to samples I and II) when
studying trends at fixed redshift. 

First, it is interesting to consider the effect of galaxy magnitude on
the blue fraction. In doing this, it is important to use the same
absolute 
magnitude cuts at all colours (\emph{i.e.},  vertical lines in
Figure~\ref{fig:colmag}).
If one instead used tilted cuts like the ones that define
samples II--IV (Equation~\ref{eqn:magcut}), the extreme slopes of
these cuts in colour-magnitude space would mean that 
the brightest magnitude bins would exclude all red galaxies 
while keeping some blue ones, leading to obviously spurious results. 
Thus we are limited here to subsamples of sample I, the
smallest sample in this study. 
It is also important to allow the magnitude bins
to evolve with redshift in the same manner as $M^\ast$, to ensure that
similar galaxies are being compared across the redshift range.  Thus, we
divide sample I into bins in the quantity $M_B-5\log{h}+Q(z-1)$, with
$Q=-1.37$.

\begin{figure}
\centering
\epsfig{width=\linewidth, file=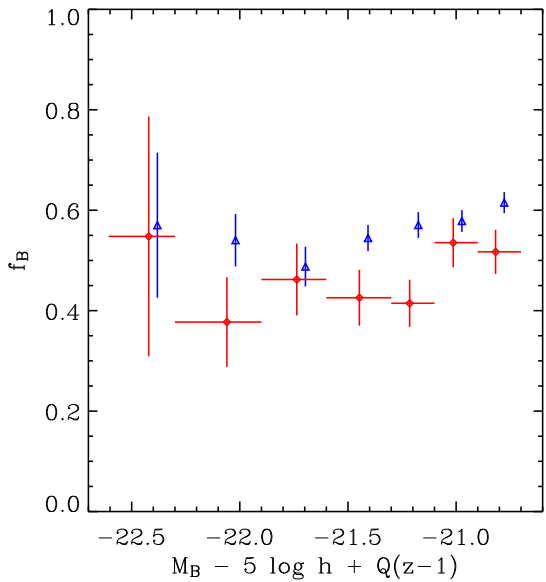}
\caption{Dependence of the blue fraction $f_b$ on absolute
magnitude in sample I ($0.75\le z \le 1.0$), for galaxies in groups
(diamonds with horizontal bars) and in 
the field (triangles). The data points show $f_b$ in bins of
evolution corrected absolute magnitude, $M_B-5\log{h}+Q(z-1.0)$, where
we have 
adopted the \citet{Faber06} value of $Q=-1.37$.  Horizontal bars
(suppressed for the field sample) show the bin size used to compute
each data point, and vertical
bars show the $1\sigma$ error on $f_b$, computed from
bootstrap resampling of the galaxies in each bin.  Data points have
been offset slightly in the horizontal direction for clarity.
Groups contain fewer blue galaxies than the field at all
magnitudes except possibly in the brightest bin.  In addition, a trend
is apparent for both group and field galaxies, 
with brighter galaxies being redder on average, except at the
bright end, where the trend may reverse. Linear fits to both
trends, excluding the brightest bin, show that their slopes are
consistent with the same value.}  
\label{fig:fb_mag}
\end{figure}

Figure~\ref{fig:fb_mag}, shows $f_b$
in bins of absolute magnitude for galaxies in groups and in the field.
Groups 
here are defined to be those systems that have two or more members in
sample I.  A trend is evident for both populations, with
brighter subsamples having a lower blue fraction---except for the
brightest galaxies, where the trend may reverse.  This last result is
consistent with the results of \citet{Cooper06a}, who find a
population of bright, blue DEEP2 galaxies in high-density
environments; it can possibly be ascribed to AGN and starburst
activity.  
It is also interesting to compare the trends for the fainter group and
field populations. 
Fitting a straight line for each subsample, excluding the brightest
bin, yields a slope of
$0.102\pm  0.033$  for the field population and $ 0.106 \pm
0.044$ for the group galaxies.  That is, there is evidence for
a similar trend with limiting magnitude for $f_b$, both in groups and
in the field.  We will discuss the implications of this result in
\S~\ref{sec:discussion}.

\begin{figure*}
\centering
\epsfig{width=0.8\linewidth, file=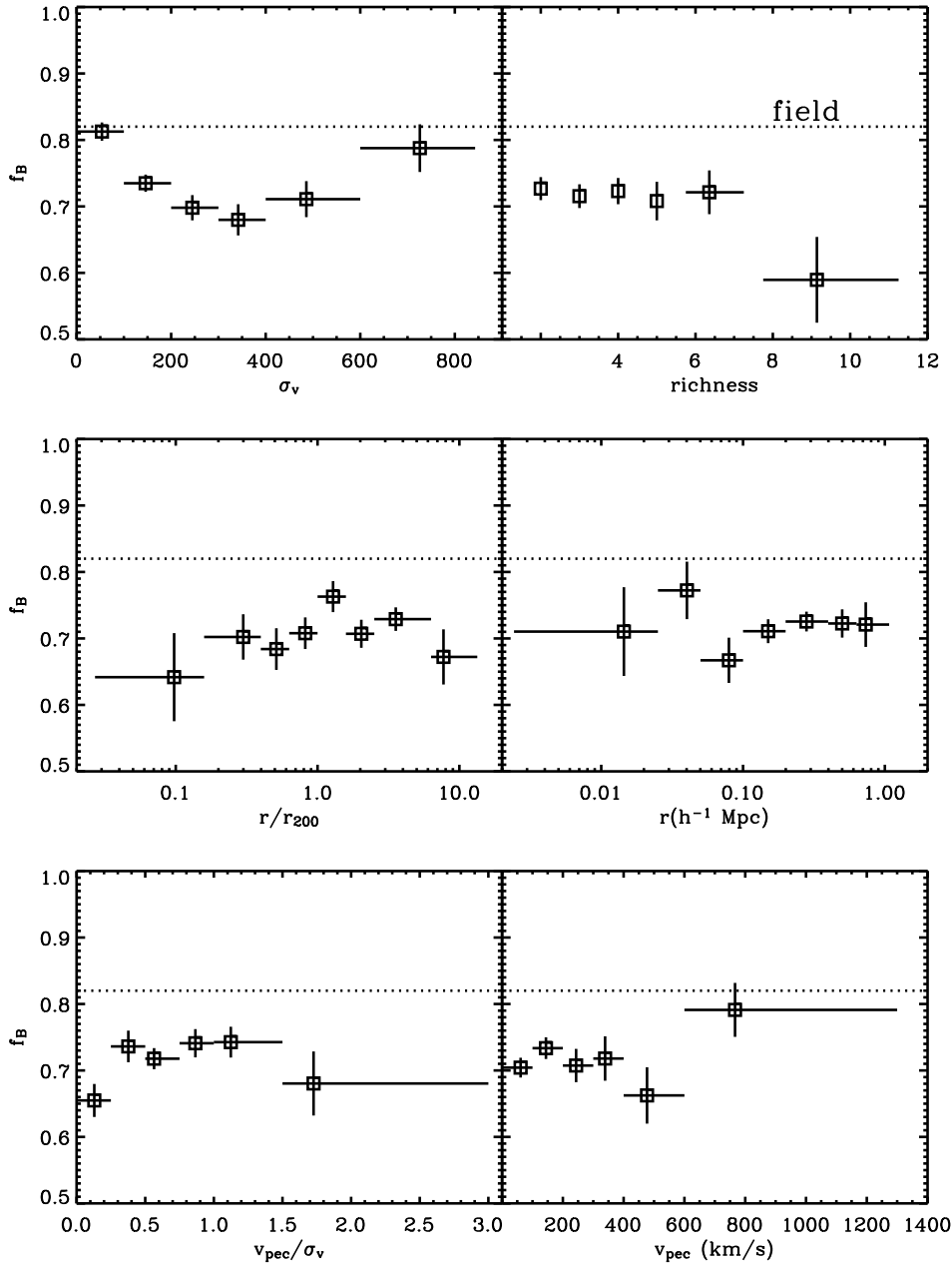}
\caption{Dependence of the blue fraction $f_b$ of group galaxies in
  sample II ($0.75\le z \le 1.0$) on various group parameters: 
  from upper left to 
  lower right, these are the velocity dispersion of the host group,
  $\sigma_v$; the richness 
  (number of galaxies above the magnitude limit) of the host group;
  distance from the centre of the host group, in
  units of $r_{200}$ (see 
  text) and in 
  megaparsecs; and peculiar velocity relative to the host group's
  redshift, in units of the host group's velocity dispersion, and in
  $\kms$. In each panel, the data points show $f_b$ in bins whose sizes
   vary to ensure robust statistics; the bin sizes are shown by
  the horizontal bars.  As in Figure~\ref{fig:fb_mag}, vertical bars
  show the $1\sigma$ bootstrap errors.%, and 
  The value of $f_b$ for field galaxies is shown as the
  dotted line. 
  See \S~\ref{sec:fb_local_trends} for
  details of the selection in each panel and
  \S~\ref{sec:computing_fb} for
  details of the computation of $f_b$.} 
\label{fig:fb_cuts}
\end{figure*}

We will use Sample II to explore other possible systematic trends in
$f_b$ at fixed $z$, since it is significantly larger than sample I by
virtue of the tilted colour-magnitude selection criterion used to
define it.  
Figure~\ref{fig:fb_cuts} shows $f_b$ for the \emph{group} galaxy
population only, binnned by various group or galaxy
properties.  For comparison, the blue fraction for field galaxies is
shown as a dotted 
line.  The upper left panel shows the dependence of $f_b$ on the 
velocity dispersion $\sigma_v$ of the galaxies' host groups---that is,
each bin contains all of the galaxies in groups with $\sigma_v$ in the
range of that bin.  In this panel, we have included in the group
sample those galaxies in groups with $\sigma_v<100 \kms$, unlike all
other plots in this paper.  As shown, in terms of $f_b$
these galaxies bear a much stronger resemblance to the field than to
the rest of the group sample.  This is no surprise, since these
groups 
are known to be predominantly false detections---\emph{i.e.}, chance
associations 
of field galaxies\footnote{it is true, though, that such
extremely poor groups in the local Universe are in fact frequently
dominated by 
spirals~\citep{ZM98}.} \citep{Gerke05}.  Figure~\ref{fig:fb_cuts} then
stands as further  
justification for our decision to class these galaxies with the
field.   The other data points in this panel show an interesting trend
with $\sigma_v$: $f_b$ declinines with $\sigma_v$ at low
dispersion and then rises again at high dispersion.  Caution is
warranted, 
however, because (1) high dispersion groups are more likely to be
contaminated by interloper field galaxies, since the VDM group finder
uses a larger search volume for such groups; and (2) the
large uncertainty in the measured values of $\sigma_v$ for DEEP2
groups makes it difficult to assess the reality of observed trends of
galaxy properties with group velocity dispersion.  In particular,
point (2) might be a problem because the
sample has many groups with two members, for which measurements of
$\sigma_v$ are maximally uncertain.  Such groups might dominate over
the population of \emph{true} high-dispersion systems, which are
rare.  If there were, say, an underlying monotonic decline in $f_b$
with increasing $\sigma_v$, then the up-scattered small groups could
induce an apparent upturn at the high-$\sigma_v$ end.  For these 
reasons, we refrain from drawing conclusions regarding the relation
between $f_b$ and $\sigma_v$.  

There is, however, an apparent trend with group richness, $N$, at high
values of $N$, visible in
the top right panel of the figure.  Richness in this context is
defined to be the
number of galaxies in a given group above the colour-magnitude limit
$M_{cut}$ that defines sample II (see Equation~\ref{eqn:magcut} and
Table~\ref{tab:samples}).  The
blue fraction declines at the highest richness values.
This result
illustrates the primary reason that we
included in our sample only those groups with two or more members above
$M_{cut}$: had we not done so, we would have been sampling richer
groups at higher redshifts, potentially causing a spurious
decrease in $f_b$ with increasing redshift.

The remaining four panels investigate the dependence of $f_b$ on group
members' distance 
from their host groups' centres (group-centric radius) and on their
peculiar velocity relative to their host groups.
To probe dependence on group-centric radius, we compute $f_b$ for
group galaxies within annuli 
on the sky around the mean right ascension and declination of their
host group.  The annuli are defined in two ways---in units of
comoving Mpc and also in units of $r_{200}$, the radius at which
the group is 200 times denser than the background.  This radius can be
estimated from the group's radial velocity dispersion $\sigma_r$ as 
\begin{equation}
r_{200}=\frac{\sqrt{3}\sigma_r}{10H(z)},
\label{eqn:r200}
\end{equation}
where $H(z)$ is the redshift-dependent Hubble
parameter \citep{CYE97}.  
We compute each group member's peculiar velocity with respect to the
mean redshift of its host group as $v_{\mathrm{pec}} =
c(z-\bar{z})/(1+\bar{z})$, and we consider the peculiar velocity both
in units of $\kms$ and normalized to the host group's velocity
dispersion $\sigma_v$.  The middle two panels of
Figure~\ref{fig:fb_cuts} show the 
dependence of $f_b$ in groups on the group-centric distance, and the
bottom two panels show the dependence on peculiar velocity.  None of
these plots shows a significant trend, and linear fits to the data
points in each panel are consistent with zero slope.

The lack of a trend with group-centric radius may be surprising in
light of studies of groups and clusters at low (\emph{e.g.},
\citealt{Whitmore93, DeP04})
and high (\emph{e.g.}, \citealt{Postman05}) redshift, showing a strong
relation between cluster-centric 
radius and galaxy type.  However, there is a large uncertainty in the
determinations of DEEP2 group centres and mean redshifts since these
involve 
taking means of a small number of objects, so it is likely that any
trends would be significantly diluted by scatter in these
values.\footnote{Indeed, the particularly attentive reader will have
  noticed that 
  the highest measured values 
of $r/r_{200}$ here are $\sim 10$, which is extremely large for any
realistic system; such values are attributable to groups with
spuriously low measured values of $\sigma_v$ and hence $r_{200}$.} 
Nevertheless, there is a hint in the lower left panel that 
galaxies at low values of $v_{\mathrm{pec}}/\sigma_v$ and $r/r_{200}$
tend to be 
redder than the general group population.  In addition, in the lower
right panel it appears that group galaxies with moderate
absolute peculiar velocities are redder than usual, while those with
the highest velocities are blue.  This echoes the trend with $\sigma_v$
at upper left (as expected, since only high-$\sigma_v$ groups contain
galaxies with high values of $v_{\mathrm{pec}}$); these two panels
taken together suggest that a negative correlation between $f_b$ and
$\sigma_v$ 
for DEEP2 groups may be masked by the presence of interloper field
galaxies that preferentially lie at high values of
$v_{\mathrm{pec}}$.   It would be preferable if we could compute
Fig.~\ref{fig:fb_cuts} using only high-richness groups, whose
dispersions and centers can be determined more robustly; however,
restricting even to groups with $N > 3$ reduces the sample size so
significantly that no firm conclusions 
can be drawn in light of the resulting error bars.  Thus, in the 
absence of any significant measured trends, we see no justification
for further restricting our sample by peculiar velocity or group-centric
radius---for instance by taking only galaxies within $r_{200}$ as has
been done in various other 
studies (\emph{e.g.}, \citealt{DeP04, Poggianti06}), in which it was
possible to measure $r_{200}$ more robustly than it is here. 
In what follows, the sample 
of group galaxies will always include \emph{all} galaxies in groups
with $\sigma_v>100 \kms$ and with two or more members above
$M_{\mathrm{cut}}$, regardless of radius or peculiar velocity.

\subsection{The evolution of the blue fraction}
\label{sec:fb_evol}

Having ensured that the group sample will not be affected by trends at
fixed redshift that will complicate comparisons at 
different redshifts, we now proceed to
probe the evolution of the blue fraction with $z$.
We divide each of the samples in Table~\ref{tab:samples} into several
independent redshift bins containing equal numbers of galaxies,  
and we
compute $f_b$ in each bin for
the group and field galaxy subsamples. 
As discussed in \S~\ref{sec:fb_errors}, there may be scatter from bin
to bin that exceeds the error estimates, especially at $z\sim 0.9$, so
to smooth out this effect  we also compute $f_b$ in a sliding box
whose width is twice the 
average width of the independent bins in each sample.

\begin{figure*}
\centering
\epsfig{width=0.8\linewidth, file=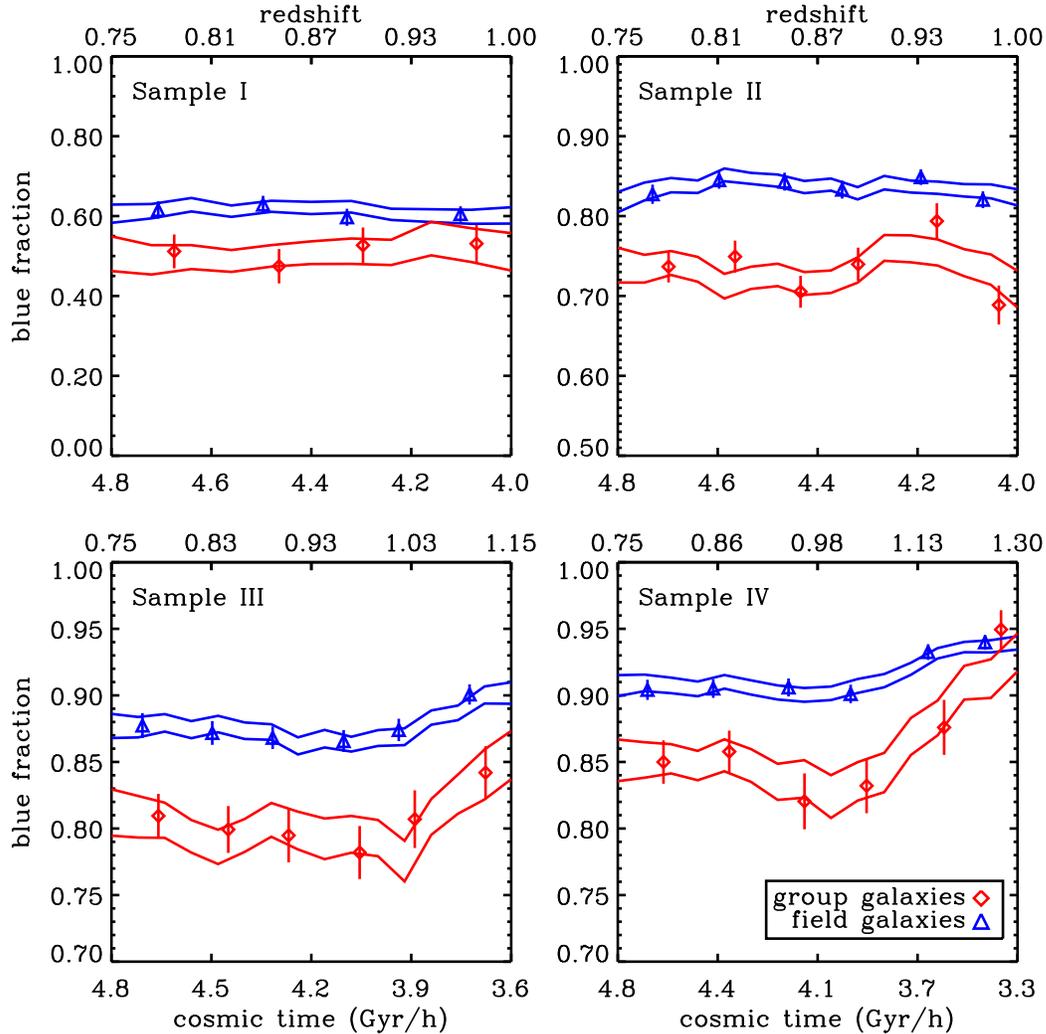}
\caption{Evolution of the blue fraction with redshift and cosmic time
  (time since the big bang).  For all four samples
  defined in   
  Table~\ref{tab:samples}, the evolution of $f_b$ is shown from
  $z=0.75$ to the 
  limiting redshift of each sample.  
%(from left to right) $z_{\mathrm{lim}}=1.0$, $1.15$ and $1.3$.  
In each 
panel, data points show $f_b$ values in six independent
bins (only four for sample I, which is much smaller than the
others) containing equal numbers of galaxies (Note that these
bins are typically broader than the ones used in
Figure~\ref{fig:colmag};
each bin contains a significant number of red galaxies).    
Blue triangles show $f_b$ for
field galaxies, and red diamonds denote $f_b$ for group galaxies; these
points have been slightly offset in the redshift direction for
clarity.  Lines 
show 1$\sigma$ errors on $f_b$ computed in a sliding bin of twice the
mean width of the independent bins.  Note that the four panels' axes
have different horizontal and vertical scales; the absence of the
upturn at high redshift in the top two panels is owing to the fact
that these panels do not extend beyond $z=1$.  Over the redshift
ranges in common, all samples are fairly consistent.  In particular,
no sample shows significant evolution in 
$f_b$ over the range $0.75<z<1$, but for $z\ga 1$ (samples III and
especially IV), $f_b$ evolves 
substantially, with much stronger evolution in groups than in the
field.  
The group and field samples become nearly
indistinguishable at the highest redshifts considered.  See
\S~\ref{sec:computing_fb} for details of the computation of the blue
fraction.} 
\label{fig:fb_evol}
\end{figure*}

The resulting $f_b$ values for
each sample are shown in Figure~\ref{fig:fb_evol} as a function of
redshift and of cosmic time; this Figure summarizes the principal
results of this paper.  
As noted previously, there is clearly a significant difference
in the blue fractions of group and field populations at $z\sim 
1$. In addition, in all four samples there is 
no evidence for evolution in $f_b$ over the range $0.75 < z < 1.0$:
group and field populations are consistent with constant
$f_b$ over this range. 
However, the bottom two panels of the Figure show that  $f_b$
evolves dramatically at $z \ga 1.0$, and this
evolution is much stronger in groups than it is in the field.  
In fact, interestingly, the two populations appear to converge in
terms of $f_b$ at $z\sim 1.3$.  It is also notable that the evolution
in samples III and IV appears to flatten below $z\sim 1$, in agreement
with samples I and II.  It is important to emphasize that the bins
used in the lower right panel of
Figure~\ref{fig:fb_evol} are broader than those shown
in Figure~\ref{fig:colmag}, so that each bin contains a substantial
number of red galaxies; that is, the convergence of the group and field
populations at $z\sim 1.3$ does not result from small-number
statistics. 

To further assess the robustness of these results, we
have investigated the effects of changing our assumptions about the
DEEP2 redshift incompleteness modeling and the luminosity evolution of
the sample.
In each panel of Figure~\ref{fig:fb_evol},  the galaxy weighting
scheme used is  the ``optimal'' scheme of \citet{Willmer06}, and the
magnitude evolution 
parameter is set to the value found in \citet{Faber06},  
$Q=-1.37$.  Changing the
weighting scheme to either the average or minimal models
of~\citet{Willmer06}---or even using no weights whatsoever---effects
no qualitative change in the results shown in
Figure~\ref{fig:fb_evol}.  Similarly, changing the value of $Q$ by
$\pm 0.31$ (the $1\sigma$ uncertainty on this parameter from
\citealt{Faber06}) has no qualitative effect on the results. 

As a further test, we also consider
the possibility of evolution in the colour of the ``green valley''
separating red and blue galaxies in colour-magnitude space.  
If the colour of the valley changes with time, the $f_b$ evolution seen
in Figure~\ref{fig:fb_evol} might simply 
be the result of passive evolution moving galaxies redward across our
(fixed) dividing line (\emph{e.g.}, \citealt{Andreon06b}).  However,
the existence of a bimodality in the 
galaxy colour distribution suggests that the division between red
and blue galaxies should be drawn through the valley, as we have drawn
it, rather than being tied to the colours of red
galaxies alone. Simple inspection of Figure~\ref{fig:colmag} shows no
clear evidence for evolution in the locus of this valley.  Indeed, if
one adopts a model in which red galaxies are continuously being formed
from blue ones throughout the redshift range of interest (\emph{e.g.},
the 
``hybrid'' model proposed by \citet{Faber06}), then one would expect 
little  evolution in the valley's position.  As old red galaxies
evolve passively toward redder colours, new red galaxies are formed to 
take their place, resulting in a red sequence that broadens with time,
while the valley between red and blue galaxies remains nearly fixed.

\begin{figure}
\centering
\epsfig{width=\linewidth, file=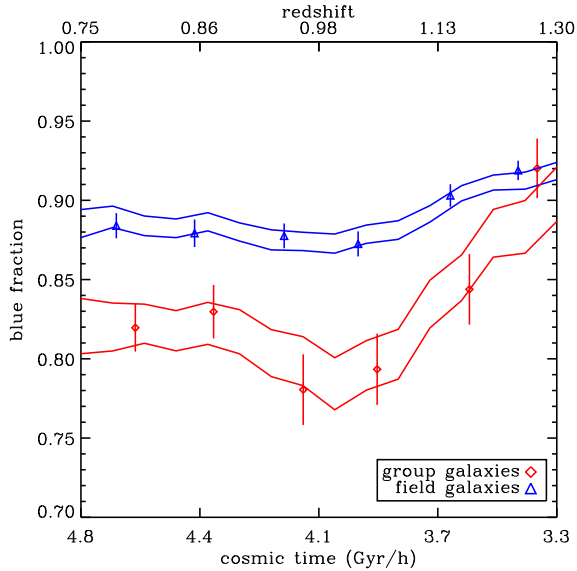}
\caption{Evolution of $f_b$ in sample IV, when the dividing line
   between red and blue galaxies is allowed to evolve redward by 0.1
  magnitudes per unit decrease in $z$ \citep{Blanton06}.  There is a
  (probably spurious) drop in $f_b$ out to $z=1.0$ (see text), but
  at higher redshifts the evolution is the same as observed with a
  non-evolving separation in Figure~\ref{fig:fb_evol}.}
\label{fig:btevol}
\end{figure}

Nevertheless, it is worth exploring the effect of allowing the
dividing line to evolve.  
\citet{Blanton06} showed that the locus of the green valley has
evolved redward by at most $\sim 0.1$ magnitudes in $u-g$ from $z=1$
to the 
present day.  We test the effect of this evolution by applying a colour
separation which is given by Equation~\ref{eqn:colourcut} at $z=1$,
with a linear evolution of 0.1 magnitudes redward per unit decrease in
redshift (that is, the blue-red dividing line gets redder with time,
as might be expected from passive evolution). 
The evolution of $f_b$ with
this evolving cut 
in sample IV is shown in Fig~\ref{fig:btevol}.   As shown, the
evolving colour separation produces a (marginally significant)
\emph{drop} in the group $f_b$ from $z=0.75$ 
to $z=1.0$, probably because the green-valley evolution we assume here
is too strong.  This is followed by a rise at higher redshifts and a
convergence of the group and field values at $z\sim 1.3$.  
Hence, the
main results of Figure~\ref{fig:fb_evol} still hold: there is 
substantial
evolution in $f_b$ in groups at $z\ga 1.0$, with a convergence of the
group and field $f_b$ values at $z\sim 1.3$. 

Before this result can be fully accepted, however, there is another
systematic effect that should be considered.   It has been observed
(\emph{e.g}, \citealt{Bell04a, Weiner05}) that there are two distinct
classes of 
red galaxies at high $z$: nonstar-forming, ``red-and-dead'' early-types
and dust-reddened, star-forming late-types.  The
latter class will 
presumably exhibit spectral emission lines; hence it should be
relatively easy to obtain redshifts for these galaxies, even if their
spectra have low signal-to-noise ratios.  Redshifts may be harder
to obtain for the nonstar-forming class, since these exhibit weaker
spectral features in general\footnote{However, it has recently been
  shown \citep{Yan06} that $\sim 40\%$ of  nonstar-forming galaxies in
  the SDSS 
  exhibit significant [O II] emission from AGN activity;
  redshifts should be readily obtainable for these galaxies.}.  It
is thus possible that DEEP2 fails to 
obtain redshifts preferentially for high-redshift 
(\emph{i.e.} faint) early-types.   This would mimic evolution in $f_b$
both in groups and in the field, but if early-type galaxies occur
preferentially in groups, this selection effect could potentially also
lead to the convergence seen in the group and field $f_b$ values at
$z\sim 1.3$.  
In principle, the weighting scheme described in
\S~\ref{sec:weights} should approximately correct for this effect;
however, if the situation is so dire that
\emph{all} galaxies at a given redshift and in a particular region of 
colour-colour-magnitude space are redshift failures, then the
weighting will be of no help. 

This potential bias must therefore be taken seriously. Fortunately, it
is possible to get some sense of its importance by looking at the
\emph{HST}/ACS imaging that the AEGIS collaboration has obtained in
the EGS~\citep{AEGIS}.  Supposing that the
true mix of spheroidal and dusty-spiral red galaxies \emph{in the
  field} is constant with redshift, if early-types are being
preferentially 
missed at high redshift, this would result in a declining ratio of
dead to 
dusty among red field galaxies in DEEP2.  We have examined all of the
red field galaxies with \emph{HST} imaging  and confirmed DEEP2 
redshifts in the range $0.75<z<1.3$ in the EGS (a total of 63
galaxies).  At all redshifts, roughly $50\%$ of these show 
no visible evidence for star formation or spiral structure.  Hence, it
appears that DEEP2 is \emph{not} preferentially failing to obtain
redshifts for a particular class of red galaxies at high $z$, although
it is worth emphasizing that this conclusion has been drawn from a
limited sample of galaxies\footnote{It may be possible to address this
  issue more directly in the near future, by using \emph{Spitzer}/IRAC
  data to obtain photometric redshift information for DEEP2 redshift
  failures.}.

\begin{figure}
\centering
\epsfig{width=\linewidth, file=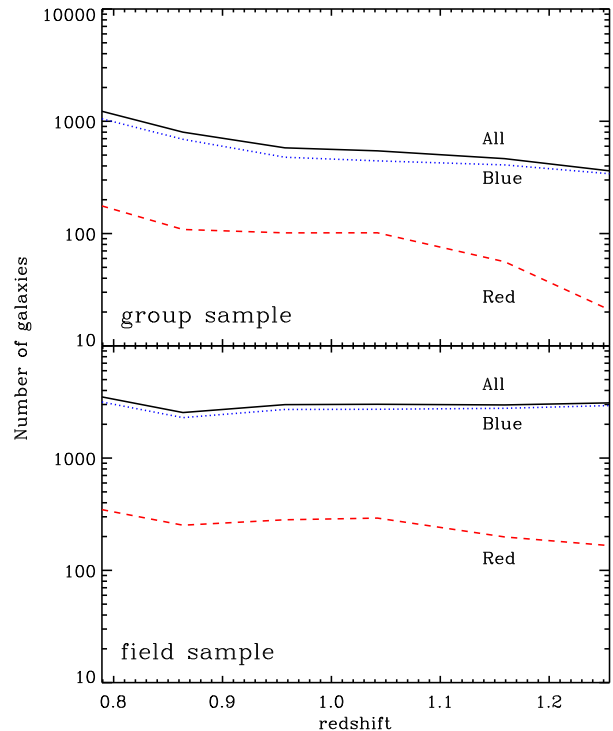}
\caption{Incompleteness-corrected galaxy counts used to compute $f_b$
  for sample IV in  Figure~\ref{fig:fb_evol}.  Solid lines show
  the total number  
  of galaxies, $\mN_t$ (see
  Equation~\ref{eqn:N_def}), in groups (top) and the field 
  (bottom), for the 
  six redshift bins shown in the bottom right panel of
  Figure~\ref{fig:fb_evol}.
  Dotted lines show the number of blue galaxies, $\mN_b$, in
  these bins, and
  dashed lines show the number of red galaxies,
  $\mN_r=\mN_t-\mN_b$.  As shown, while the
  overall number of group galaxies is growing with time, the number
  of red galaxies in groups increases much more rapidly,
  increasing by almost an order of magnitude over the redshift range
  shown.} 
\label{fig:bo_numbers}
\end{figure}

The strong evolution of the DEEP2 group blue fraction at $z\ga 1.0$
therefore appears to be robust.  It
can be explained 
either by a blue galaxy population that decreases with time, a red
galaxy population that increases with time, or both. As shown in
\citet{Bell04b}, \citet{Willmer06} and \citet{Faber06}, there is
significant  
growth in the typical comoving number density $\phi^\ast$ of
red-sequence galaxies over the redshift range probed by DEEP2, while
$\phi^\ast$ for blue galaxies has remained roughly constant over this
range.  Hence, it seems reasonable to ascribe the evolution of
$f_b$ to the preferential build-up of the red galaxy population in
groups.  This 
conclusion is made manifestly clear in
Figure~\ref{fig:bo_numbers}.  Rather than showing fractions, this
figure shows the (incompleteness-weighted) numbers of galaxies from
sample IV in each of the redshift bins shown in the lower right panel of
Figure~\ref{fig:fb_evol}, both in groups and in the field.  More
specifically, the figure shows the total number of galaxies
$\mN_t$, the number of blue galaxies $\mN_b$ (see
Equation~\ref{eqn:N_def}), and the number of red
galaxies $\mN_r=\mN_t-\mN_b$, all as a function of
redshift, both in groups and in the field.  It is important to reiterate
that the redshift bins were chosen to contain equal \emph{unweighted}
total numbers of galaxies, so the overall
weighted numbers in each bin are not constant.  In any case, we find
that the results of this paper are insensitive to the weighting scheme
we choose and remain unchanged even if no weights are applied at all.

The figure shows that $\mN_t$ for groups increases with time, as
expected in the hierarchical $\Lambda$CDM structure-formation
paradigm.  But $\mN_r$ in groups grows at a much faster rate,
increasing by almost an order of magnitude from $z\sim1.3$ down to
$z\sim0.75$.  At the same time $\mN_t$ in the field stays nearly
constant, and $\mN_r$ in the field increases only modestly (by a
factor of $\la 2$) over this range.  It seems clear, then,
that the build-up of red sequence galaxies over the DEEP2 redshift
range has taken place preferentially within groups and clusters.

\section{Tests with mock catalogues}
\label{sec:mock_tests}
The above results so far appear to be robust, but to establish them
definitively it is vital that we also test for biases introduced by the
group-finding procedure.  For example, it is well known (and
has been confirmed here) that groups and clusters are populated by
redder galaxies than the field, but, as shown in
Figure~\ref{fig:colmag}, the DEEP2 apparent magnitude limit 
selects strongly against red galaxies at high redshift.  If red
galaxies dominate groups and clusters, this effect
might cause their
richnesses to appear strongly reduced at high redshift---so much so
that they might eventually become undetectable with the VDM
group-finder.  The high-redshift VDM group sample would then be
dominated by false detections. 
If this were the case, then the group and field populations
identified by the VDM would necessarily be indistinguishable, since
they would essentially be random subsamples of the same galaxy
population, and so the convergence of the group and field
populations seen in Figure~\ref{fig:fb_evol} would simply be the result
of group-finding errors and not of genuine galaxy evolution in
groups.  

\begin{figure}
\centering
\epsfig{width=\linewidth, file=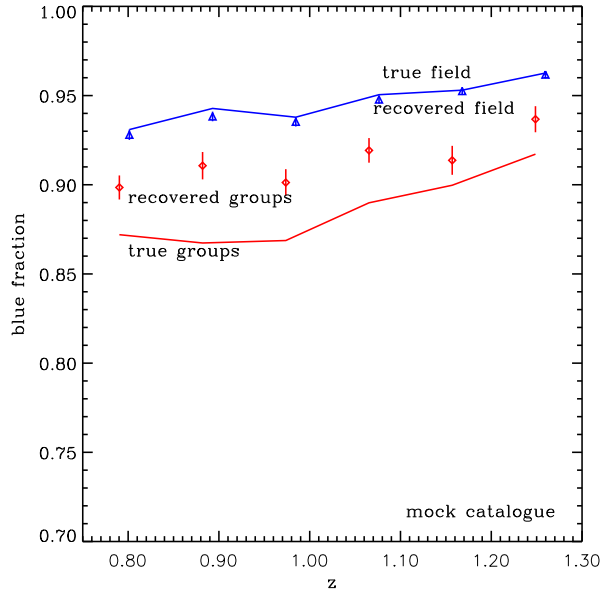}
\caption{Evolution of the blue fraction in the mock catalogues described
in Section~\ref{sec:mocks}. 
Solid lines show the \emph{true} evolution
of $f_b$ in the mocks, in groups (lower curve) and in the field (upper
curve), computed using \emph{real} 
groups, \emph{i.e.}, sets of galaxies that actually occupy common dark
matter haloes.  %%The lower panel shows 
Data points show the \emph{reconstructed} evolution of
$f_b$ in the mocks, computed using \emph{found} groups, i.e., groups
of galaxies 
identified in redshift space by the VDM group-finding algorithm.  As
expected, the 
distinction between found groups and the field is smaller than the
distinction for real groups because of interlopers; however, it is
clear that the group-finder does \emph{not} induce a reconstructed
$f_b$ evolution that is stronger than the actual evolution in the
mocks or as strong as what is observed in the data.} 
\label{fig:bofoundreal}
\end{figure}

We can test for such an effect by running the VDM group-finder on
the DEEP2 mock catalogues described in \S~\ref{sec:mocks}.  
As discussed in that section, the mock
galaxies are assigned rest-frame $U-B$ colours according to the
observed DEEP2 
colour-magnitude-environment in the redshift range
$0.8<z<1.0$.  Figure~\ref{fig:bofoundreal} shows the evolution
of $f_b$ in the mocks for galaxies in
\emph{real} groups (\emph{i.e.}, galaxies that 
actually reside in dark matter haloes that contain other galaxies,
lower curve), for 
galaxies in \emph{found} groups (i.e., galaxies in groups identified
by the VDM algorithm, diamond points), and for the field samples that
complement each 
group sample (upper curve and triangle points).  In computing $f_b$,
the same limit in colour-magnitude 
space is applied as the one used for sample IV
(given by Equation~\ref{eqn:magcut} and Table~\ref{tab:samples} and
shown in 
Figure~\ref{fig:colmag}), and, as in the data, galaxies are included
in the group sample only if their host group has two galaxies brigher
than the limit.  As Figure~\ref{fig:bofoundreal} illustrates, the VDM 
group-finder reduces the separation in $f_b$ between the group and
field 
samples (as expected, since the found groups are
contaminated by interloper field galaxies).  However, it 
does not induce a spurious evolutionary trend in the blue fraction of
the found group sample: the slopes of $f_b$ versus $z$ are consistent
for the real and found samples.  Thus it does not appear that the VDM
group finder is 
introducing spurious trends in the measurement of blue fraction
evolution. 

Finally, it is interesting to note in Figure~\ref{fig:bofoundreal}
that the values of $f_b$ for the \emph{true} group and field samples
in the mocks
show clear evolution.  As discussed in Section~\ref{sec:mocks}, we have
not introduced any redshift dependence in the colour-environment
relation used to assign colours to mock galaxies. By design, the blue
fraction will remain constant with redshift for comparable
local density values.  Therefore, the 
evolution of $f_b$     
seen in the mocks can only be due to evolution in the
\emph{distribution of local densities} within the mocks.  That is, the
growth of 
large-scale structure in the mocks causes the number of high-density
regions to grow with time, so there are
fewer red galaxies at high redshifts in the mocks simply because there
are fewer galaxies in overdense environments at high redshifts.  This
effect provides a partial explanation for the evolution in $f_b$
observed in DEEP2. However, because the mock group and field samples
remain distinguishable in terms of $f_b$ at all redshifts while the
observed samples do not, there must also be
some decrease in the blue fraction of DEEP2 group
galaxies with time at \emph{fixed} overdensity. (This can also be seen
directly in Figure~\ref{fig:bo_numbers}, where the numbers of red
galaxies in 
groups grow more rapidly than the total group population.) We will
discuss this 
further in Section~\ref{sec:discussion}.

%%%%%%%%%%%%%%%%%%%%%%%%%  SECTION 4 %%%%%%%%%%%%%%%%%%%%%%%%%%%
\section{Discussion}
\label{sec:discussion}

The main results of this paper are (1) a
significantly lower value of $f_b$ in group versus field
populations at $z\sim 1$; (2) a
strong increase of $f_b$ in groups with redshift at $z\ga 1.0$; (3)
the \emph{lack} 
of any obvious $f_b$ evolution, for either groups or the field, at
lower  redshifts; (3) a
convergence of $f_b$ for the group and field populations at $z\sim
1.3$; (5) a decline in 
$f_b$ in richer groups;
and (6) a negative correlation 
between $f_b$ and  galaxy luminosity, both in groups and in the 
field. In light of
Figures~\ref{fig:fb_evol} and \ref{fig:bo_numbers}, it appears
especially clear that the build-up of the DEEP2 red 
galaxy population has taken place most dramatically in groups and
clusters rather than in the field (here and throughout, \emph{field}
refers to all galaxies not in groups).  

In interpreting this conclusion, three interrelated questions
arise.  
\begin{enumerate}
\item What physical mechanisms are acting in groups and clusters to
  quench star formation and produce the evolution seen in
  \S~\ref{sec:fb_evol}? 
\item Is the evolution of $f_b$ in DEEP2 groups and clusters
  simply an extension of the similar evolution seen at low
  redshift---\emph{i.e.}, is this just the BO effect at high
  $z$, or are other mechanisms in effect at this epoch?
\item If groups and clusters are the site of red galaxy formation,
  what is the nature of the red galaxies present in the field,
  especially those at high redshift?
\end{enumerate}
By comparing these results to other studies at high and low
redshift---and to current theoretical models of galaxy evolution---we
can attempt to fit our 
observations into a coherent picture of star formation and quenching
in groups and clusters.

\subsection{Comparison to the evolving colour-density relation}
\label{sec:environcompare}

As shown in \S~\ref{sec:fb_evol}, the blue fraction in DEEP2
groups evolves strongly over the range $1.0\la z \la 1.3$, and the
group and field populations become indistinguishable in terms of $f_b$
at $z\sim 1.3$.  These results are qualitatively consistent with
recent work studying the evolution of the high-redshift galaxy
population as a function of local galaxy density, both in DEEP2 and in
other surveys.  In a companion paper to this one, \citet{Cooper06b}
show that a correlation exists between galaxy red fraction
($f_r=1-f_b$) and local overdensity out to $z>1$ but that this
correlation vanishes at $z\sim 1.3$, in close agreement with our
findings here.  

\citet{Nuijten05}
used galaxies in the photometric CFHT Legacy Survey to show that the
fractions of red and morphologically early-type 
galaxies grow with time and evolve more strongly in high-density
regions than elsewhere.  In addition, \citet{Tanaka05} compared
photometric observations of two high-redshift clusters to the SDSS
galaxy population; they find that the color-density relation has
steepened with time and that this steepening is strongest in very
overdense regions. 
However, as shown in \citet{Cooper05}, photometrically determined
redshifts are not sufficient for accurate determinations of galaxy
density.  Also, \citet{Nuijten05} consider galaxies above a set of 
\emph{fixed} absolute magnitude limits rather than allowing their
limits to evolve as $M^\ast$. Since galaxies that are faint relative
to $M^\ast$ tend to be bluer than average (see Figure~\ref{fig:fb_mag}),
and because $M^\ast$ is brighter back in time, a fixed cut in $M$
includes more faint (blue) galaxies at high redshift, which causes
$f_b$ to increase (spuriously) with z.  This selection effect acts to
\emph{enhance} the observed evolution.

Also, \citet{Cucciati06} recently considered the fractions of galaxies
in various colour ranges as a function of redshift and local density in
the VVDS redshift survey.  They find
that the density dependence of the blue and red fractions vanishes at
$z\sim 1.0$, a somewhat lower redshift than we find (see
\citealt{Cooper06b} for a detailed discussion).
They also find that the blue fraction falls 
with time at \emph{all} densities, in contrast to our finding
that $f_b$ is roughly constant in the field at all redshifts.  We see
are two possible reasons for this (minor) disagreement: first, these
authors 
again consider only a fixed absolute magnitude limit rather than
taking $M^\ast$ evolution into account, and second, as they also
caution, their faint, red galaxy sample is incomplete at the highest
redshifts 
they consider, so any evolution of $f_b$ in this range will be
spurious.  In this paper, we have been careful to design galaxy
samples that avoid such subtle selection effects.  We are thus able to 
confirm that the results from the three papers above \citep{Nuijten05,
  Tanaka05, Cucciati06} are
qualitatively correct
despite possibly suffering from such effects.   

We have also considered 
groups versus the field, rather than local density.  This will
be useful in that it allows our results to be interpreted in
terms of the dark matter haloes hosting DEEP2 galaxies and compared
more readily to theory (see Section~\ref{sec:sams}).  Local
density remains a powerful tool for studying galaxy populations,
however, since it probes a wider range of environments than groups and
clusters, allowing one to explore whether galaxy evolution might be
driven by processes that occur on scales larger (or smaller) than the
scale of groups and clusters, or whether \emph{underdense} regions are
as important to galaxy evolution as overdense regions.
We therefore urge the reader to compare our results to those of
\citet{Cooper06b}.  In particular, that paper shows that the red
fraction of galaxies in overdense regions declines strongly with
increasing redshift, while remaining constant in underdense regions,
and that the red fractions of these two environments converge at $z\sim
1.3$.  This result is strikingly consistent with the lower right panel
of Figure~\ref{fig:fb_evol} and stands as confirmation that the results
presented in this paper are correct.

Both the overdensity considerations of \citet{Cooper06b} and the
halo-mass considerations of this paper point to a partial answer to the
question of which mechanisms are driving the evolution we observe.  
As discussed in \S~\ref{sec:groupcat}, DEEP2 samples intermediate-mass 
objects (in the range $5\times 10^{12}\la M/M_\odot \la 10^{14}$), rather
than rich ($M\sim 10^{15} M_\odot$) clusters.  Also, 
\citet{Cooper06b} find that the relation between galaxy colour and
overdensity is a smooth function of overdensity and does not set in
only at the highest densities. We may thus
confidently  
conclude that the quenching of star formation in groups and clusters
cannot be  ascribed \emph{solely} to processes that are significant
only in rich clusters, such as ram-pressure
stripping and harassment (similar conclusions have been drawn in
lower-redshift studies as well, \emph{e.g.}, \citealt{Balogh02}).   

\subsection{Comparison to the Butcher-Oemler effect}

It is nevertheless tempting to identify the $f_b$ evolution seen in
Figure~\ref{fig:fb_evol} with a simple extension of the 
BO effect\footnote{Recall that, in this paper, we are using the phrase
``Butcher-Oemler effect'' to refer specifically to evolution of the
blue fraction in groups and clusters owing to the formation of
large-scale structure and the attendant lower age of typical massive
haloes back in time (see \S~\ref{sec:intro}).} to higher redshifts.  It is
difficult, however, to square this 
interpretation with the weak or nonexistent evolution seen over the
range $0.75\le z \la 1.0$.  To be sure, this
redshift range might not sample a long enough period of cosmic time to
discern evolution due to the BO effect: ``classical'' detections of
the BO effect (\emph{e.g.}, \citealt{BO84, RS95, MD00, KB01}) have
typically considered clusters over a redshift range 
$0\la z \la 0.4$, which corresponds to $3 h^{-1}$ Gyr of
cosmic 
time, whereas the range $0.75\le z \le 1$ covers only $0.8h^{-1}$ Gyr.
Nevertheless, at minimum, if the evolution observed here is simply
the BO effect extended to high-redshift groups, the evolution becomes 
substantially more rapid at $z\ga 1.0$.  

Moreover, it is not even clear that groups of the sort being
considered in this study would be expected to exhibit a strong BO
effect at lower redshifts.  As shown in Figure~\ref{fig:samplehists}, the
vast majority of DEEP2 groups have velocity dispersions $\sigma_v \la
600 \kms$.  Because most groups also have very few galaxies with
spectroscopy 
(typically fewer than ten), there is also a very large scatter
(typically a few hundred $\kms$) in the measured dispersions.  Coupled
with the fact that the 
number of groups falls very steeply with increasing $\sigma_v$, this
scatter implies that the measured dispersions will tend to be biased
high by an Eddington-type bias, so that DEEP2 groups are typically not
as massive as their velocity dispersions seem to imply\footnote{Note
  that this point does \emph{not} affect the estimate that DEEP2
  groups lie in the range $5\times10^{12}\la M/M_\odot \la 10^{14}$;
  this was derived from the known masses of group haloes in the mock 
  catalogues.}.   Also, as
discussed above, our mock catalogues imply that DEEP2 
does not sample high-mass clusters.  There have 
been very few studies of galaxy  
evolution in the intermediate-mass systems we \emph{do} sample, but a
recent study by 
\citet{Poggianti06} has shown that the fraction of [O II] emitting
galaxies does not evolve significantly out to $z\sim 0.8$ for groups
with $\sigma_v \la 500 \kms$.  Also, in their study of the
morphology-density relation at $z\sim 1$, \citet{Smith05} show that 
galaxies in intermediate-density environments show little
morphological evolution down to the $z\sim 0.5$ epoch studied by
\citet{Dressler97}.  If we naively identify intermediate-density
environments with intermediate-mass groups like the ones in DEEP2, we
might also not expect to find significant evolution in 
$f_b$ over the range $0< z \la 1$.  Indeed, preliminary indications
(Gerke et al. in prep.) are that $f_b$ is little different in DEEP2
groups out to  
$z\sim 1$ than it is in the groups detected by \citet{Eke04} in the
2dFGRS, at least for the galaxy population considered in sample I (but
see \citealt{MOL06}).   

In light of the theoretical basis for the BO effect, it is not
surprising that the effect would appear weak in the DEEP2 
sample.  As first discussed by \citet{Kauffmann95a}, the BO effect is a
natural outcome of hierarchical structure formation so long as
cluster environments are efficient at quenching star formation in
galaxies.    Since clusters form at late times, galaxies
in clusters observed at high redshift will have spent less time on
average in the cluster environment than their counterparts observed
at low redshift.  Hence, they will have had
less time to feel the effects of the physical mechanisms 
that are acting to quench star formation.  This picture
implies 
that the BO effect will be weaker in less massive systems, as is
observed in, \emph{e.g.} \citet{Poggianti06}, since 
smaller systems will typically form earlier and then merge into larger
systems, resulting a similarly young galaxy population at all
redshifts in these objects.  \citet{KB01}  
considered a set of simple models for the star-formation histories of
cluster galaxies, and their results support the basic
\citet{Kauffmann95a} scheme: the 
observed BO effect can be explained by the decreasing accretion rate
onto clusters at late times, coupled with a declining star-formation
rate in field galaxies.  Neither of these studies
considered the effect of dark energy on this picture, but it is worth
noting that, in the concordance $\Lambda$CDM 
cosmology, the universe becomes vacuum dominated around $z\sim
0.4$, sharply attenuating matter infall onto clusters at later times.
Infall rates 
should be much higher at the redshifts considered in this paper; hence
one might expect a weaker BO effect in DEEP2, since the quenching blue
galaxy 
population in groups and clusters is being steadily replaced by new
infalling galaxies.

It therefore seems reasonable to conclude that the evolution seen at
$z\ga 1$ in
DEEP2 is different in nature from the low-redshift BO
effect.  This is 
not to say that the physical mechanisms quenching star formation in
\emph{individual galaxies} are necessarily different at the two
epochs; rather, the evolution may simply be in the \emph{efficiency}
with which these mechanisms alter the bulk properties of the cluster
galaxy population.  In particular, the rapid evolution at $z\ga 1.0$
and the convergence of the group
and field $f_b$ values at $z\sim 1.3$ suggest a picture in which, at
the epochs considered in this paper, 
groups have only recently become efficient environments for
quenching, while the group galaxy population has a similar
star-formation rate to the field at earlier times.  For example, if
the quenching 
processes in groups cause star formation to end on a time scale
$\tau_Q \sim 1$ Gyr (which best reproduces the BO effect in
\citealt{KB01}), then DEEP2 groups in this picture would have begun
quenching their member galaxies efficiently only 1 Gyr prior to the
earliest epoch considered here, \emph{i.e.}, at a redshift $z\sim
1.7$ (for $h=0.7$).   

This basic conclusion (that the quenching
mechanism turns on only at $z\la 2$) is consistent with observations
of high star-formation rates in massive galaxies at $z\sim 2$
(\emph{e.g.}, \citep{Daddi04}). In addition, it is supported by
the stellar-population study of \citet{Schiavon06}, which shows that
nonstar-forming galaxies in DEEP2 at 
$z\sim 0.9$ have mean stellar ages of roughly $1.5$ Gyr (indicative of
quenching at 
$z\la 2$) \citep{Schiavon06}.  Moreover \citet{Harker06} have shown
that a relatively simple model of stellar population evolution, in
which star 
formation is turned off starting at $z\sim 2$, can explain the build-up of
the red sequence from $z\sim 1$ to $z\sim 0$.    
The conclusion that DEEP2 groups only started quenching their member
galaxies at 
$z\sim 2$ is also nicely consistent with the emerging theoretical
picture of galaxy formation (\emph{e.g.}, \citealt{DB06}), in
which star formation is quenched efficiently only in haloes with masses 
above $\sim 10^{12} M_\odot$ at $z\la 2$.   We discuss this last point
in more detail in \S~\ref{sec:sams} below. 

It is reasonable to worry, though, that the contamination of the
group population by 
interloper field galaxies might cause the two populations to appear
indistinguishable when, in truth, a tiny difference still exists.
Certainly Figure~\ref{fig:bofoundreal} shows that such contamination
weakens the observed difference between the two populations (although
it does not induce spurious evolutionary trends).  However,
the tests on mock catalogues in \citet{Gerke05} show that the sample of
DEEP2 galaxies in VDM groups is dominated by galaxies that are truly
in groups, so it should be possible to discern all but the tiniest
differences between the samples.  In addition, recall that
\citet{Cooper06b} find a similar convergence in the \emph{red}
fraction of galaxies in the most 
underdense and most overdense environments.  Since that study does not
suffer from the same contamination effects as the VDM group catalogue,
we may be confident that the apparent convergence of the two
populations' blue fractions cannot be
ascribed solely to interloper contamination.

\subsection{Comparison to semi-analytic models of galaxy formation} 
\label{sec:sams}
It will be instructive to briefly compare the basic picture that the
observations in \S~\ref{sec:results} imply---namely, that groups are
not efficient at quenching 
star formation at $z\sim 2$---with the picture that emerges from
semi-analytic models (SAMs) of galaxy formation.  This comparison is
especially worthwhile because most SAMs have been designed to
reproduce the details of the galaxy population at $z\sim 0$, so
comparisons to the high-$z$ DEEP2 population will provide sharp tests
of their correctness.  We will limit ourselves here to qualitative
descriptions and arguments, however, and defer detailed comparisons to
future work.

In making the comparison, we will sometimes use as a concrete example
the model used by \citet{Croton06} to populate the Millenium run
\citep{Millenium}, a very large, high-resolution N-body simulation of
cosmic structure formation.  The elements of this model are
qualitatively very similar to other recent SAMs \citep{Bower06,
  Cattaneo06, KJS06}, all of which have been successful at
reproducing the observed 
luminosity functions and colour distributions of galaxies at $z\sim
0$. 
In the \citet{Croton06} model, dark matter haloes are populated with
two types of galaxy \citep{Springel01}: each halo contains one
\emph{central} galaxy and may also incorporate
\emph{satellite} galaxies, whose parent haloes
merged with the main halo at some time in the past (this conceptual
division of the galaxy population is nearly universal in modern-day
SAMs).  
Each halo is also assumed to
contain a baryonic component consistent with the overall cosmic baryon
fraction.  This gas can fuel ongoing star formation if it cools and is
accreted 
by the halo's central galaxy (\emph{e.g.}, \citealt{WR78, WF91});
hence, in order to cease star formation and 
permanently join the red sequence, a galaxy must stop accreting new cold
gas onto its disk.  

Most SAMs admit two basic routes by which gas accretion can be 
stopped, associated with the two types of galaxy in the halo.   In the
case of satellite galaxies, their parent
haloes' gas is simply added to the gas reservoir of the larger halo
when they are accreted.  Gas is then only allowed
to accrete onto its central galaxy (which sits at the minimum of the
potential 
well), so model satellite galaxies will cease star formation forever
once they 
have exhausted the gas supply in their discs.  All
satellite galaxies will thus eventually migrate to the red
sequence via strangulation.  
For central galaxies, the situation is somewhat more 
complicated.  In all haloes, infalling gas is shock-heated to the halo
viral temperature, but, in low-mass haloes, that hot gas cools on very
short 
timescales so that accretion onto the central galaxy is very rapid.
Star formation never ceases permanently in such 
galaxies.  In more massive haloes, the cooling time is longer, and 
the gas forms a quasi-static, hot halo of gas. 
The seperation of haloes into these two regimes---the ``rapid cooling'' 
and ``static hot halo'' regimes---has a long history (\emph{e.g.}, 
\citealt{RO76, BFPR}); recent simulations show that the regimes are 
divided by a rather sharp 
transition mass of $\sim 3 \times 10^{11} M_\odot$ \citep{BD03,
  Keres04}.   

Regardless of the halo mass, however, hot gas near the centre can
still cool, accrete onto the central object, and form stars, unless
some 
other source of energy is available to stop the cooling; this is the
well-known cooling-flow problem.  \citet{Croton06} propose to solve
this problem via heating from low-luminosity AGN lurking in the haloes'
massive central galaxies.  They refer to this heating as ``radio
mode'' AGN feedback, to distinguish it from winds created by bright
AGN in the ``quasar mode'' (which should only temporarily eliminate
cold gas from a galaxy). In essence, if the central supermassive
black hole (SMBH) is massive enough, its average AGN energy output
will be sufficient to offset cooling in the halo gas and stanch the
flow of gas onto the central galaxy; this basic idea has also been
implemented in other SAMs \citep{Bower06, Cattaneo06, KJS06}.  Since
halo mass and SMBH mass 
are linked (\emph{e.g.}, \citealt{HR04}), this model effectively
predicts a threshold mass above 
which accretion onto the central galaxy ceases; then, once the galaxy 
has exhausted its existing supply of cold gas, star formation ends.

The
threshold mass above which ``radio mode'' heating can stop cooling
flows is a few times $10^{12} M_\odot$
(\citealt{DB06, Cattaneo06}; Croton et al. in prep.), roughly an order
of magnitude smaller than 
the typical masses of DEEP2 groups. Thus, the Croton et al. picture
implies 
that \emph{all} galaxies in DEEP2 groups should eventually be quenched
and join the red sequence.  The only group members that will be blue
(i.e., star-forming 
or recently star-forming) in that model will be those that recently
fell in to their 
host group and have not yet exhausted the cold gas in their discs.  
If this picture is accurate, then the convergence of $f_b$ for the
group and field populations shown in Figure~\ref{fig:fb_evol}
implies that a typical DEEP2 group member at $z=1.3$ was \emph{not} in
a halo above the quenching mass until $\tau_Q \sim 1$ Gyr before that epoch.
This is broadly consistent with hierarchical structure growth in a
$\Lambda$CDM universe.  In the Millenium Run, for example, one can
trace the history of a typical
DEEP2-group-like halo, with a mass of $\sim 10^{13} M_\odot$ at $z=1.3$.
On average, such a halo's most massive progenitor at $z\sim 2$ had a
mass of $\sim 5\times 10^{12} M_\odot$---roughly the mass at which
radio-mode heating becomes efficient.  There are $1.1 h^{-1}$ Gyr of
cosmic time between $z=2$ and $z=1.3$, a reasonable timescale for
quenching and migration to the red sequence.  
It is worth pointing out 
that this explanation is conceptually quite different than the BO 
effect:  as we have used the phrase in this paper, the BO effect
is simply a \emph{consequence} of the assumption that groups and
clusters quench star formation (and thus it affects only infalling
satellites) whereas the mechanism described here is the \emph{onset}
of quenching in group-size haloes (which affects \emph{central}
galaxies as well)\footnote{Note, however, that this effect, on its own, will
\emph{not} be 
sufficient explain the growing number density of red galaxies
over the entire DEEP2 redshift range.  The $f_b$ evolution seen in 
Figure~\ref{fig:fb_evol} appears to \emph{flatten} at $z\la 1$, so for
the red sequence to continue its build-up, there must be an increase
in the number density of groups.  That is, the growth of the red
galaxy population at $z\la 1$ might still be attributable solely to the
hierarchical growth of structure.}.

As we have seen, evolutionary results fit nicely, in a broad sense,
with the 
current semi-analytic picture of galaxy formation. However, there are
some difficulties in detail, especially as regards satellite
galaxies.  Satellites in the model are unable to accrete new gas; such
accretion is reserved for central galaxies, \emph{regardless} of the
parent-halo mass.  Under these assumptions, groups at $z\sim 1.3$
should have a lower blue fraction than the field, unless \emph{all} of
their galaxies were isolated at $z\ga 2$.  Since this seems unlikely,
it may be necessary to revise the 
assumption that \emph{all} satellites are stripped of their halo gas
with equal efficiency, regardless of host-halo mass.  
Indeed, there is already significant evidence that this is
an oversimplification.  The \citet{Croton06} model significantly
over-predicts the 
abundance of faint red galaxies (see their Figure 11). 
Also, there is evidence that a satellite galaxy's colour is correlated 
with the colour of its central galaxy \citep{Weinmann06}; this would not
occur if all satellites were stripped of their gas with equal
efficiency.  Most recently,  \citet{Weinmann06b} showed that the
Croton et al. model severely underpredicts the blue fraction
of SDSS satellites and argued that the efficiency with which gas is
stripped from satellite galaxies must scale with the mass of the host
halo.  We note in passing that such a revision to the model would
also account naturally for the correlation seen in
Figure~\ref{fig:fb_cuts} between $f_b$ and group richness, since more
massive groups would quench star formation more efficiently.

\subsection{On the nature of the red field galaxies}

We close this section by commenting on the continued presence of a
significant red galaxy population in the field at high redshift, even
when the group and field blue fractions have converged.  One of our
main conclusions in this work is that groups are the primary locus of 
quenching and red-galaxy formation, and the most naive interpretation
of this statement 
would suggest that there should be \emph{no} red
galaxies in the field.  This is an oversimplification,
of course, not only because isolated ellipticals are known to exist
locally (\emph{e.g.}, \citealt{CMZ01}), but also since the red
sequence will be contaminated by some 
fraction of dusty star-forming galaxies (\emph{e.g.} \citealt{Bell04a,
  Weiner05}).
One might also expect such galaxies 
to be more common at high redshift, when the average
star-formation rate of the Universe was higher.  It should therefore
come as no 
surprise that DEEP2 contains a non-negligible population of 
red field galaxies at all $z$, but it is not immediately clear whether 
dusty star-formers can account for all such galaxies.
 
To address this issue, we make use of the \emph{HST}/ACS imaging that
has been obtained for a portion the Extended
Groth Strip subregion of the DEEP2 survey~\citep{AEGIS}.  In this
region there are eleven red field galaxies 
in the redshift range $1.15<z<1.3$.  Five of these show spiral
structure or 
other visible evidence of star-forming activity, while six appear to
be nonstar-forming spheroidal or lenticular objects.  Thus it appears
that a significant population of early-type field galaxies remains at
these redshifts; the red field population observed in DEEP2 cannot be
fully accounted for by dusty star-formers.  

The issue is further complicated by group-finding
errors.  The VDM group-finder fails to identify some fraction of group
members; these will appear in the field sample, and some of them will
be early-types.  However, the statistics given in ~\citet{Gerke05}
imply that only $11\%$ of the field galaxies identified in sample IV
are actually group members (not the value of $6\%$ quoted in
\S~\ref{sec:groupcat}, since we reclassified a significant number of 
group galaxies as field galaxies in \S~\ref{sec:selection}).  As shown
in Table~\ref{tab:samples}, $12\%$ of these---or $1.3\%$ of the total
field sample---will be red.  Combining this with the observed fraction
of dusty star-formers still does not account for the full population
of red field galaxies (about $7\%$ of the field in sample IV).

It is therefore worth considering a few mechanisms by which \emph{bona
  fide} early-type galaxies could enter the field population.  Here
  the distinction between central and satellite 
galaxies will again be useful.  The threshold mass for quenching in
\emph{central} galaxies, a few times $10^{12} M_\odot$, is somewhat
lower 
than the mass range of groups considered in this study; therefore,
there should be a population of quenched central galaxies that do not
have satellites that enter the DEEP2 sample; these would appear as
red galaxies in the field.   This picture naturally
leads to the correlation between $f_b$ and 
limiting magnitude seen in Figure~\ref{fig:fb_mag}.  The strength of
this correlation appears to be the same both in groups and in the
field.   But if all galaxies
above a given mass are quenched, and if halo mass correlates roughly
with luminosity, then brighter subsets of the galaxy population will 
naturally have lower blue fractions, both in groups and in the field.
This explanation also fits nicely with the recent results
of~\citet{Conroy06}, who show that the brightest galaxies'
mass-to-light ratios have grown dramatically between $z=1$ and the
present day while
$M/L$ has remained roughly constant for fainter galaxies---presumably
  because the brightest galaxies have ceased 
star formation but continue to accrete new, non-luminous mass. 

In addition, it has
been shown that winds from merger-driven quasar activity may expel
cold gas from galactic disks \citep{SDH05, Hopkins05c}. If such a wind
arises in a 
galaxy whose halo is less massive than the threshold at which
``radio-mode'' AGN feedback is important, then the galaxy should
eventually reincorporate some of the expelled gas and resume star
formation.  But if the timescale for reincorporation is longer than
the lifetime of hot, blue stars ($\sim 1$ Gyr), then such galaxies
will temporarily 
join the red sequence.  Such temporarily red galaxies could also
account for some red field objects.

It is also noteworthy that the blue fraction in the field
appears to evolve at $z\ga 1$, if less strongly than in groups. 
This can be seen in the lower right panel of Figure~\ref{fig:fb_evol}:
$f_b$ in the field changes by $\sim 0.05$ between $z=1.0$ and $z=1.3$,
while 
the group value changes by $\sim 0.1$.  It is worth exploring
whether 
this effect is real, or whether it can simply be attributed to
contamination of the field sample by group galaxies, arising from
group-finding errors.  Let us assume, to start, that the \emph{true}
blue fraction in the field is
constant with $z$. Then, if the field sample 
contains a fraction $c_f$ 
of contaminating group galaxies, and the group sample similarly has a 
contamination fraction $c_g$ of field galaxies, it is
straightforward to show that, for a given observed change $\Delta
f_b^\mathrm{group}$ in the group blue fraction, the
observed change in the 
field blue fraction is 
\begin{equation} 
\Delta f_b^\mathrm{field} = 
\frac{c_f}{1-c_g} \Delta f_b^\mathrm{group}.
\label{eqn:contam_evol}
\end{equation}
For sample IV, the success statistics given in \citet{Gerke05} imply
that $c_g=0.46$ and $c_f=0.11$ (as discussed earlier in this Section).
Thus, 
the observed change in the field blue fraction should be $\sim 20$\% 
of the observed change in the group blue fraction.  This ratio is, in
fact, 
marginally consistent with the error bars in Figure~\ref{fig:fb_evol}.
Hence, the observed decline in the field blue fraction with time is
largely caused by 
contamination by misclassified group galaxies.  This conclusion is
supported by the findings of \citet{Cooper06b}, who observe no
evolution in the red fraction of galaxies in the most underdense
environments.  On the basis of the present study alone, however, it
remains possible that there is also some small amount of  intrinsic
evolution in the field population.
Such evolution would not be too surprising: since the characteristic
halo mass $M^\ast$ does not reach the quenching mass of a few times
$10^{12}M_\odot$ until $z\sim 1$
(\emph{e.g.}, 
\citealt{Reed03}), the number density of quenched central galaxies in
the field will be growing relatively rapidly at earlier times.
Regardless, it 
remains clear from Figures~\ref{fig:fb_evol} and \ref{fig:bo_numbers}
that the build-up of the red sequence in DEEP2 takes place
preferentially in groups.     

%%%%%%%%%%%%%%%%%%%%%  CONCLUSION %%%%%%%%%%%%%%%%%%%%%%%%%%%%
\section{Conclusion}
\label{sec:conclusion}

We have used the DEEP2 group catalogue of \citet{Gerke05} to study the
blue fraction of galaxies in groups and the field at $z\sim 1$.  After
creating four samples that are carefully designed to be free from
colour-dependent and redshift-dependent selection effects, we probe the
dependence of $f_b$ on a galaxy's environment (group or field), on
galaxy luminosity, on group properties, and on
redshift.  

First, over the redshift range $0.75\le z <1$, there is a
significant difference in $f_b$ between 
the DEEP2 group and field populations, with a substantially lower
$f_b$ in groups than in the field. The
well-known colour segregation observed between nearby group and field
galaxy populations is thus already in place at $z\sim 1$.   In
addition, there is a negative correlation between $f_b$ and group
richness over 
this redshift range, echoing local correlations between galaxy colour
and various proxies for halo mass.  A trend also exists at these
redshifts between $f_b$ and the luminosity of the galaxies being
considered, with fainter galaxies being more likely to be blue; the
strength of this trend does not appear to depend on 
whether those galaxies are in groups or the field.

No siginificant evolution in $f_b$ occurs in the DEEP2 group or field 
galaxy populations over the range
$0.75\le z \la 1$; however, at $z\ga 1$, there is dramatic,
statistically significant evolution in
the group population, with $f_b$ rising rapidly as $z$ increases.
Most notably, the value of $f_b$ in groups becomes indistinguishable from
the field value at $z\sim 1.3$.  To assess the robustness of these results, 
we have constructed mock catalogues that replicate the colour-environment
correlations in the DEEP2 data.  Tests with these mock catalogues
indicate that the evolution and convergence are not artifacts of the
group-finding algorithm.  Moreover, these tests suggest that the
observed evolution can be explained partially---but not entirely---by 
evolution in the distribution of galaxy environments, \emph{i.e.} by
the growth of large-scale structure, coupled with the known
correlation between environment and galaxy properties.  However, to
explain the strong evolution and convergence, some further
mechanism is required.

Because the $f_b$ evolution appears to weaken at $z\la 1$, we
conclude that it is different in nature from the classical
BO effect, in which $f_b$ in clusters rises dramatically 
from $z=0$ out to $z\sim 0.5$.  Indeed, the BO effect is a natural
outcome of hierarchical structure formation, \emph{provided that}
groups and clusters are already efficient at quenching star formation
in the 
galaxies that fall into them.  But there is a \emph{convergence}
between  $f_b$ values in groups and in the field at $z\sim 1.3$, which
implies that DEEP2 groups were, at some epoch, \emph{not} especially
efficient at 
quenching infalling galaxies.  In particular, if it takes $\sim 1$ Gyr
for a quenched galaxy to join the red sequence, our results indicate
that the groups in this study only became suitable environments for
quenching at $z\sim 2$.

This conclusion is broadly consistent with current semi-analytic
models of galaxy formation \citep{Croton06, Bower06, Cattaneo06,
  KJS06}. 
In those models, the central galaxies of dark matter haloes are quenched
when low-level AGN feedback becomes strong enough to stop gas in the
halo from cooling and condensing onto the galaxy to form stars; this
process becomes efficient at a threshold halo mass of a few times
$10^{12} M_\odot$ (\citealt{DB06, Cattaneo06}; Croton et al in prep.).
Our mock catalogues imply that the groups observed in DEEP2
have masses in the range $5\times 10^{12} M_\odot \la M \la 5\times
10^{13} M_\odot$.  A typical halo in this mass range at $z\sim 1.3$
will drop below the threshold mass for quenching $\sim 1h^{-1}$ Gyr
earlier, at $z\sim 2$.  This
is a potential explanation for the convergence in the group and field
$f_b$ values at $z\sim 1.3$.  Present-day SAMs may have
difficulty explaining these observations in detail, however,
particularly because they universally assume that \emph{satellite}
galaxies are 
quenched in all haloes regardless of mass or redshift.  In the future,
it will be worthwhile to make detailed comparisons between our
observations and the precise predictions of SAMs, in order to
refine the physical description of galaxy formation that they
provide.

%\acknowledgments
\section*{Acknowledgments}
We thank Christian Marinoni for making his original VDM group-finding
code available for our use, and we thank Guinevere Kauffmann,
Michael Boylan-Kolchin, and especially Peder Norberg for 
valuable discussions.  This work was supported in part by NSF grants 
AST00-71048, AST00-71198, AST0507483 and AST0507428.  JAN is
supported by Hubble Fellowship 
HST-HF-01165.01-A.  SMF would like to thank the Miller Institute at
UC-Berkeley for  the support of a visiting Miller Professorship during
part of this work.  The data presented herein were obtained at the
W.M. Keck Observatory, which is operated as a scientific partnership
among the California Institute of Technology, the University of
California, and the National Aeronautics and Space Administration. The
Observatory was made possible by the generous financial support of the
W.M. Keck Foundation.  The DEIMOS spectrograph was funded by a grant
from CARA (Keck Observatory), an NSF Facilities and Infrastructure
grant (AST92-2540), the Center for 
Particle Astrophysics, and by gifts from Sun Microsystems and the Quantum
Corporation.  The DEEP2 Redshift Survey has been made possible through 
the dedicated efforts of the DEIMOS staff at UCO/Lick Observatory, who
built 
the instrument, and the Keck Observatory staff, who have supported it on
the telescope.  Finally, the authors wish to recognize and acknowledge
the very significant cultural role and reverence that the summit of
Mauna Kea has always had within the indigenous Hawaiian community.  We
are most fortunate to have the opportunity to conduct observations
from this mountain.

%\bibliography{../bibs.bib} 

%\pagebreak

%%%%%%%%%%%%%%%%%%%%%%%%% FIGURES %%%%%%%%%%%%%%%%%%%%%%%%%%%%%%

\end{document}